\documentclass[fleqn,usenatbib]{mnras}

\usepackage{newtxtext,newtxmath}

\usepackage{amssymb}	

\usepackage[T1]{fontenc}

\DeclareRobustCommand{\VAN}[3]{#2}
\let\VANthebibliography\thebibliography
\def\thebibliography{\DeclareRobustCommand{\VAN}[3]{##3}\VANthebibliography}

\usepackage{graphicx}	
\usepackage{amsmath}	
\usepackage{pdflscape}
\usepackage{threeparttable}

\newcommand{\hii}{H\,{\sc ii}}

\title[ESO~137-001 Kinematics]{Tracing the kinematics of the 
whole ram pressure stripped tails in ESO~137-001}

\author[Rongxin Luo et al.]
{Rongxin Luo$^{1}$\thanks{E-mail: rongxinluo217@gmail.com},
Ming Sun$^{1}$\thanks{E-mail: ms0071@uah.edu},
Pavel J\'{a}chym$^{2}$,
Will Waldron$^{1}$,
Matteo Fossati$^{3,4}$,
Michele Fumagalli$^{3,5}$,
\newauthor
Alessandro Boselli$^{6}$,
Francoise Combes$^{7}$,
Jeffrey D. P. Kenney$^{8}$,
Yuan Li$^{9}$,
Max Gronke$^{10}$
\\
$^{1}$Department of Physics, University of Alabama in Huntsville, 
301 Sparkman Dr NW, Huntsville, AL 35899, USA\\
$^{2}$Astronomical Institute of the Czech Academy of Sciences, Bo\v{c}n\'{i} II 1401, 141 00, Prague, Czech Republic\\
$^{3}$Dipartimento di Fisica G. Occhialini, Universit\`{a} degli Studi di Milano Bicocca, Piazza della Scienza 3, I-20126 Milano, Italy\\
$^{4}$INAF-Osservatorio Astronomico di Brera, via Brera 28, I-20121 Milano, Italy\\
$^{5}$INAF-Osservatorio Astronomico di Trieste, Via G. B. Tiepolo 11, I--34143 Trieste, Italy\\
$^{6}$Aix Marseille Univ, CNRS, CNES, LAM, Marseille, F-13013 France \\
$^{7}$Observatoire de Paris, LERMA, College de France, CNRS, PSL Univ, Sorbonne University, UPMC F-75014 Paris, France\\
$^{8}$Yale University Astronomy Department, P.O. Box 208101, New Haven, CT 06520-8101 USA\\
$^{9}$Department of Physics, University of North Texas, Denton, TX 76203, USA\\
$^{10}$Max Planck Institut für Astrophysik, Karl-Schwarzschild-Straße 1, D-85748 Garching bei München, Germany
}

\date{Accepted XXX. Received YYY; in original form ZZZ}

\pubyear{2023}

\begin{document}
\label{firstpage}
\pagerange{\pageref{firstpage}--\pageref{lastpage}}
\maketitle

\begin{abstract}
Ram pressure stripping (RPS) is an important process to affect the evolution of cluster galaxies 
and their surrounding environment. We present a large \textit{MUSE} mosaic for ESO~137-001 and its 
stripped tails, and study the detailed distributions and kinematics of the ionized gas and stars. 
The warm, ionized gas is detected to at least 87 kpc from the galaxy and splits into three tails. 
There is a clear velocity gradient roughly perpendicular to the stripping direction, which decreases 
along the tails and disappears beyond $\sim45$ kpc downstream. The velocity dispersion of the ionized 
gas increases to $\sim80$ km s$^{-1}$ at $\sim20$ kpc downstream and stays flat beyond. The stars 
in the galaxy disc present a regular rotation motion, while the ionized gas is already disturbed 
by the ram pressure. Based on the observed velocity gradient, we construct the velocity model for 
the residual galactic rotation in the tails and discuss the origin and implication of its fading 
with distance. By comparing with theoretical studies, we interpreted the increased velocity dispersion 
as the result of the oscillations induced by the gas flows in the galaxy wake, which may imply an 
enhanced degree of turbulence there. We also compare the kinematic properties of the ionized gas 
and molecular gas from {\em ALMA}, which shows they are co-moving and kinematically mixed through 
the tails. Our study demonstrates the great potential of spatially resolved spectroscopy in probing 
the detailed kinematic properties of the stripped gas, which can provide important information for 
future simulations of RPS.
\end{abstract}




\section{Introduction}
Galaxy clusters can accelerate the evolution of galaxies within them. The dense environment leads to gas deficiency 
in the cluster galaxies and drives these galaxies to be redder, more spheroidal, and less star forming than their 
counterparts in the field (e.g., \citealt{Dressler1980,Lewis2002,Boselli2006}). Ram pressure stripping (RPS) is an 
important hydrodynamical mechanism responsible for the environmental effects of cluster 
galaxies evolution (e.g., \citealt{Gunn1972,Quilis2000,Boselli2022}).
As galaxies move in the hot ($T \sim 10^7$ - $10^8$ K, $n_{\rm ICM} \sim 10^{-4}$ - $10^{-2}$ cm$^{-3}$) intracluster 
medium (ICM) (e.g., \citealt{Sarazin1986}), the drag force from the ICM exerts the ram pressure on them. This pressure 
not only reorganizes the distribution and kinematic state of the interstellar medium (ISM), but also strips it out 
from the galaxies, which can rapidly quench the star formation activity in a short time-scale ($\lesssim$ 1 Gyr) 
(e.g., \citealt{Roediger2007,Boselli2008,Tonnesen2009,Fossati2018}). A short-lived starburst can also occur in 
the galaxies during the RPS process, especially when the stripped ISM moves across the galaxy discs and creates 
instabilities/turbulence to help the collapse of the molecular clouds 
(e.g., \citealt{Gavazzi1995,Bekki2003,Lee2020,Boselli2021}).

Besides its significant role in galaxy evolution, RPS can also influence the surrounding environment of galaxies through 
the stripped ISM. After being removed from the galaxies, the stripped ISM is able to get mixed with the ambient hot ICM. 
The cold gas can be heated, excited, and evaporated, which changes the phase of the stripped ISM and induce it to evolve 
as part of the ICM (e.g., \citealt{Sun2021}, see \citealt{Boselli2022} for a review). Many of the ICM clumps are likely 
originated from the stripped cold gas, suggesting RPS can contribute to the inhomogeneity or clumpiness of the ICM 
(e.g., \citealt{Vazza2013,Ge2021a}). The stripped ISM can potentially enrich the ICM and modify its metallicity 
distribution, although the contribution is not significant as those from other mechanisms, such as the stellar 
evolution in the cluster galaxies, the stellar and AGN feedbacks 
(e.g., \citealt{Schindler2005,Domainko2006,Kapferer2007}). In addition, it is also known that 
some stripped ISM can form giant molecular clouds and turn into new stars, especially in the high-pressure 
environments, which may contribute to the intracluster light (see \citealt{Boselli2022} for a review).

RPS events have been discovered in many nearby cluster galaxies, based on the multi-wavelength observations from 
radio, mm, IR, optical, UV, and X-ray (see \citealt{Boselli2022} for a review).
These studies present the important roles of RPS in gas removing and star formation quenching/triggering 
and highlight its effects on the evolution of cluster galaxies. In addition, these observations also reveal the multi-phase 
gas content in some spectacular stripped tails, which provides a great opportunity to study the evolution of stripped ISM 
and its mixing with the ICM. 

ESO~137-001 is one of the nearest galaxies which are undergoing extreme RPS events and present clear multi-phase gas 
content in the stripped tails \citep{Sun2006,Sun2007,Sun2010,Sivanandam2010,Jachym2014}. It is a late-type spiral galaxy 
in the Norma cluster (A3627; $R_{\rm A}=2$ Mpc, $M_{\rm dyn}\sim1\times10^{15}$ M$_{\sun}$, $\sigma=925$ km s$^{-1}$) 
and is located at a projected distance of $\sim200$ kpc from the central cluster galaxy. Its line-of-sight (LOS) velocity 
(4680 $\pm$ 71 km s$^{-1}$; \citealt{Woudt2004}) is close to the average LOS velocity of the cluster 
(4871 $\pm$ 54 km s$^{-1}$; \citealt{Woudt2008}), 
suggesting this galaxy mainly moves on the plane of the sky. Based on the semi-analytic modeling of the possible 
orbits for ESO~137-001, \citet{Jachym2014} suggested this galaxy is currently moving with a high velocity 
of $\sim3000$ km s$^{-1}$) and is located about 100 Myr before pericenter. The RPS tails of ESO~137-001 
are first discovered with X-ray imaging from \textit{Chandra} and \textit{XMM-Newton} \citep{Sun2006,Sun2010}. 
The X-ray tails are stripped away from the cluster centre and extend to $\sim80$ kpc projected distance from the galaxy. 
There are two narrow branches in the stripped tails, which contain a total mass of X-ray gas $\sim10^{9}$ M$_{\sun}$. 
Following observations with optical imaging and spectroscopy reveal the H$\alpha$ emission from the bright \hii{} 
regions and diffuse ionized gas in the stripped tails, which extend to $\sim40$ kpc projected distance from the 
galaxy \citep{Sun2007,Sun2010}. Warm molecular gas was also detected to $\sim20$ kpc along the stripped tails of 
ESO~137-001, with a total mass more than $10^{7}$ M$_{\sun}$ \citep{Sivanandam2010}. Based on the observations 
of \textit{APEX} and \textit{ALMA}, \citet{Jachym2014,Jachym2019} detected the CO(2-1) emission to $\sim60$ kpc 
along the stripped tails and measured a total mass of cold molecular gas more than $\sim10^{9}$ M$_{\sun}$.

Wide-field optical integral-field spectroscopy (IFS) has great potential to study the RPS process, since it can provide 
rich information about the detailed physical properties and kinematic structures of the ionized gas (i.e., warm phase) 
in the stripped tails 
(e.g., \citealt{Merluzzi2013,Fumagalli2014,Fossati2016,Fossati2019,Consolandi2017,Poggianti2017,Boselli2018,Liu2021}), 
which is a key to explore the phase changing and evolution of the stripped ISM. 
\citet{Fumagalli2014} and \citet{Fossati2016} performed the \textit{MUSE} observations to cover ESO~137-001 
and the front part of its primary tail. \citet{Fumagalli2014} found the stripped ionized gas maintains the rotation 
imprint of the galaxy disc to $\sim20$ kpc downstream that was first reported by \cite{Sun2010}, and presents an 
enhancement of the velocity dispersion along the tails. \citet{Fossati2016} distinguished the diffuse ionized gas 
from the bright \hii{} regions in the stripped tails and studied its excitation mechanism. They found the 
photoionization and shock ionization models cannot explain the high values of the \mbox{[N\,\textsc{ii}]}/H$\alpha$ 
and \mbox{[O\,\textsc{i}]}/H$\alpha$ observed in the diffuse ionized gas, suggesting other mechanisms 
(e.g., thermal conduction and magnetohydrodynamic waves, etc) may play a role in its excitation. 
As discussed above, ESO~137-001 has become the RPS galaxy with the richest amount of multi-wavelength data.

To further investigate the RPS process in ESO~137-001, we combine all the archival \textit{MUSE} data with our 
own observations to complete the full coverage of the galaxy and its whole stripped tails. The map of H$\alpha$ 
surface brightness has been used in comparisons between the H$\alpha$ and X-ray emission from the diffuse gas in 
a large sample of RPS tails \citep{Sun2021}.
This research reveals the connection between the warm and hot phases in the stripped ISM and provides evidence 
of its mixing with the ICM. In this paper, we focus on the kinematic properties of the ionized gas and study their 
variations in ESO~137-001 and its stripped tails. Details about the \textit{MUSE} observations and data processing 
are described in Section \ref{sec:observations and data processing}. The kinematic maps of the ionized gas and the 
corresponding analyses of kernel density estimation are presented in Section \ref{sec:results}. In addition, 
we also compare the motions of gas and stars in the galaxy disc and explore how the gas kinematics change along 
the stripped tails. Finally, discussion and conclusions follow in Section \ref{sec:discussion} and \ref{sec:summary}. 
Throughout this paper, we adopt the same distance and scale as used in \citet{Sun2010}, a luminosity distance of 
69.6 Mpc and 1$\arcsec\ = 0.327$ kpc. As described in Section~\ref{sec:observations and data processing}, 
ESO~137-001's system velocity is fixed at 4647 km s$^{-1}$ from its central stellar spectrum.

\begin{figure*}
	\centering
	\includegraphics[width=1\linewidth]{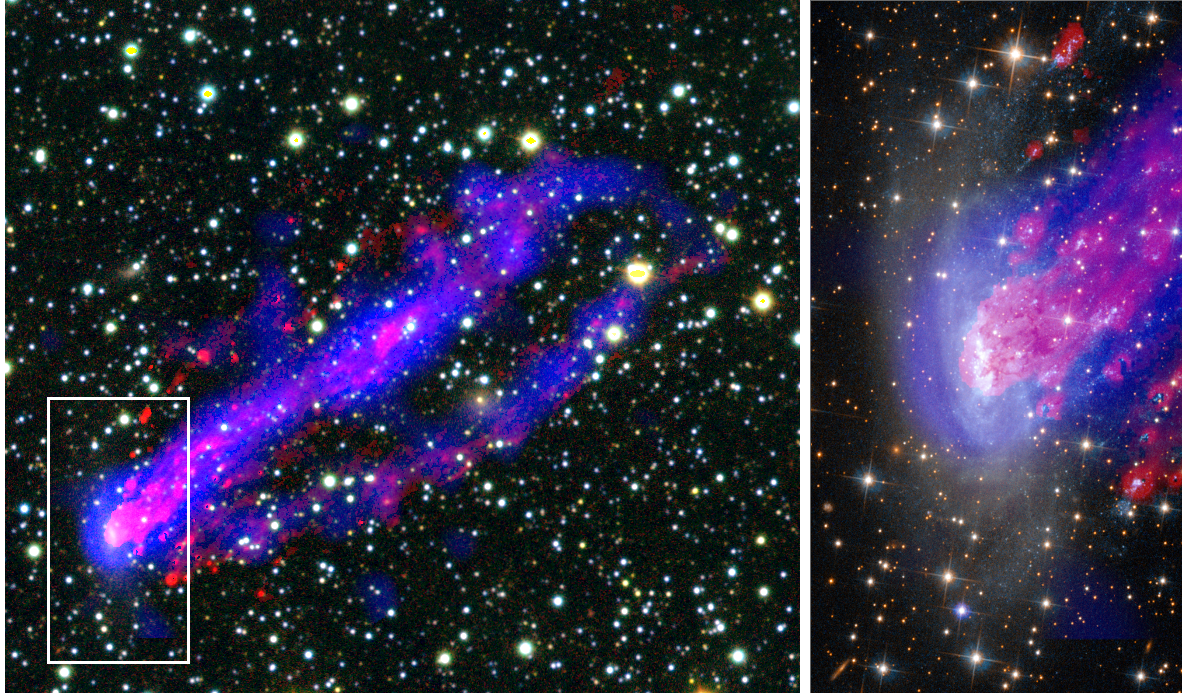}
	\vspace{-0.5cm}
	\caption{The composite images of ESO~137-001 and its stripped tails. Left panel: H$\alpha$ tails 
        (red, from this work) and X-ray tails (blue, from {\em Chandra}, \citealt{Sun2010}) overlaid on 
        the composite background of R, J, and Ks band images. The R band image is from the 
        {\em CTIO V. M. Blanco} 4-m telescope, while the J and Ks band images are from {\em VIRCAM}. 
        The white box shows the footprint of the {\em HST} image on the right. Right panel: {\em HST} 
        image of the galaxy disc overlaid with the H$\alpha$ and X-ray tails 
        (Credit of the {\em HST} image: ESA/Hubble\protect\footnotemark).}
        \vspace{-0.3cm}
	\label{fig:eso137_001_rgb}
\end{figure*}

\begin{table}
    \begin{center}
        \caption{Properties of ESO 137-001}
        \label{tab:Galaxy properties}
        \begin{tabular}{c|c}
            \hline
            \hline
            Heliocentric velocity (\text{km s$^{-1}$})$^\mathrm{a}$    & 4647 (-224)  \\
            Offset (kpc)$^\mathrm{b}$     & 180     \\
            Position Angle            & $\sim$ 9$^\circ{}$ \\
            Inclination & $\sim$ 66$^\circ{}$ \\
            M$_{\star}$ ($10^{9} M_{\odot})^\mathrm{c}$         & 5-8        \\
            L$_{\rm FIR}$ ($10^{9} L_{\odot})$$^\mathrm{d}$   &  5.2 \\
            $M_{\rm mol}$ ($10^9 M_\odot$)$^\mathrm{e}$ & $\sim$ 1.1 \\
            SFR (M$_{\odot}$ yr$^{-1}$) (Galaxy) & 1.2 \\
            Tail length (kpc)$^\mathrm{f}$ & 80 - 87 (X-ray/H$\alpha$) \\
            \hline
        \end{tabular}
    \end{center}

    Note: \\
    Adopted from \citet{Waldron2022} to summarise the key properties of ESO~137-001 relative to this work.\\
    $^\mathrm{(a)}$ The heliocentric velocity of the galaxy is determined from the stellar spectrum around 
    the nucleus (Section~\ref{sec:observations and data processing}). The velocity value in parentheses is 
    the radial velocity relative to that of Abell~3627 \citep{Woudt2004}. \\
    $^\mathrm{(b)}$ The projected offset of the galaxy from the X-ray centre of A3627.\\
    $^\mathrm{(c)}$ The total stellar mass estimated from \cite{Sun2010}.\\
    $^\mathrm{(d)}$ The total FIR luminosity from the \emph{Herschel} data (see details in \citealt{Waldron2022}).\\
    $^\mathrm{(e)}$ The total mass of the molecular gas in the galaxy from \cite{Jachym2014}.\\
    $^\mathrm{(f)}$ The full tail length.
    \label{tab:parameter}
\end{table}

\begin{figure*}
	\centering
	\includegraphics[width=1\linewidth]{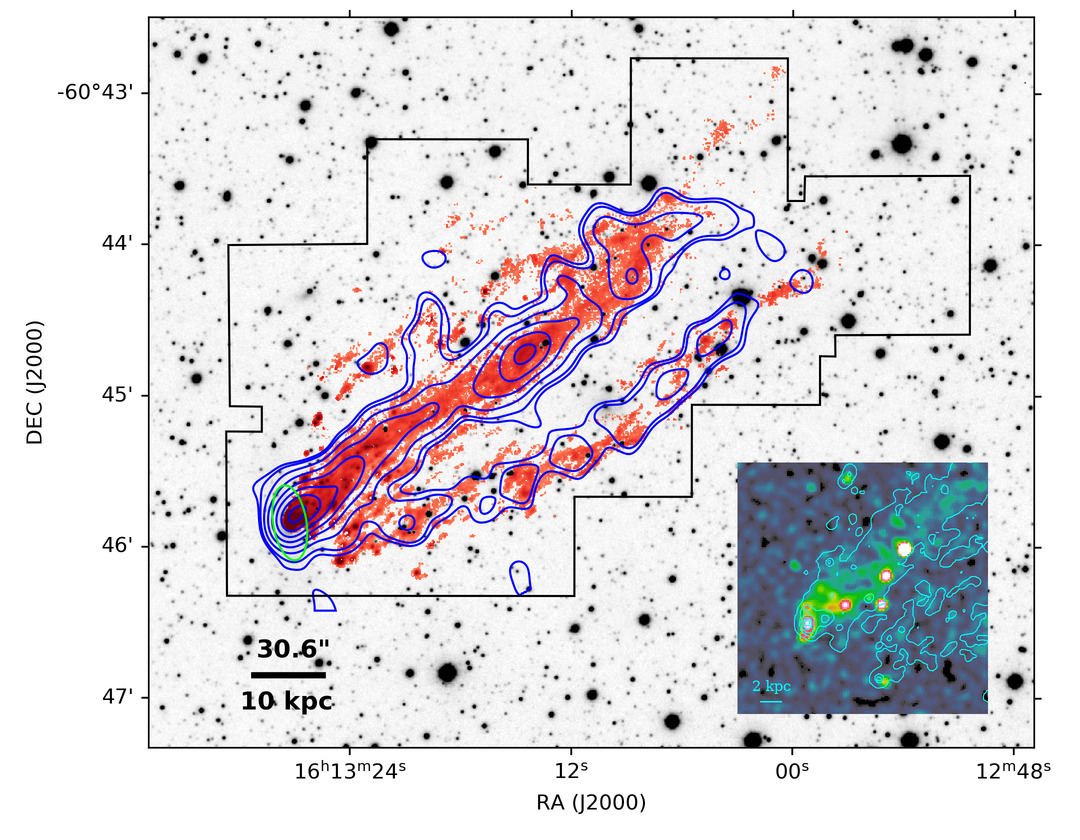}
	\vspace{-1.0cm}
	\caption{The J band image of ESO~137-001 and its stripped tails from {\em VIRCAM}. The FOV of 11 {\em MUSE} 
		pointings is enclosed in the black lines and the detected H$\alpha$ emission is overlaid in red. The contours 
		in blue show the X-ray emission observed by {\em Chandra} (\citealt{Sun2010}). The green ellipse shows the 
		half-light radius of the galaxy from the {\em HST} F160W image. One can see the good positional correlation 
		between the diffuse X-ray and H$\alpha$ emission. The X-ray stripping front and the H$\alpha$ stripping 
		front are at the same position, as shown in the zoom-in inset (0.6 - 2 keV X-ray image + cyan H$\alpha$ contours).
        Note that the blue X-ray contours are from heavily smoothed X-ray image to enhance the faint, diffuse 
        X-ray tail (see Fig. 1 of \citealt{Sun2010}) so the X-ray contours appear more extended than the H$\alpha$ 
        emission but the X-ray - H$\alpha$ spatial correlation is well established in \citet{Sun2021}.}
        \vspace{-0.3cm}
	\label{fig:eso137_001_obs}
\end{figure*}

\section{Observations and data processing}
\label{sec:observations and data processing}
We performed the \textit{MUSE} (Multi-Unit Spectroscopic Explorer, \citealt{Bacon2010}) observations for ESO~137-001 
with a wide-field mode through the European Southern Observatory (ESO) programs 095.A-0512(A) (PI: M. Sun) and 0104.A-0226(A) 
(PI: M. Sun). In addition, we also collect all the \textit{MUSE} archive data from the ESO programs 60.A-9349(A) 
(Science Verification) and 60.A-9100(G) (Science Verification). The observational dates of these programs are from 
June 21, 2014 to March 17, 2020. The seeing changes from $0\farcs57$ to $1\farcs94$ with a median of $1\farcs04$ 
and the airmass ranges from 1.24 to 1.60 with a median of 1.25. The total exposure time is 5 hours and 35 minutes. 
The detailed observational information is listed in Table~\ref{tab:observational info.}. As shown in Fig.~\ref{fig:eso137_001_obs}, 
we have a mosaic of 11 \textit{MUSE} pointings to cover the main body of ESO~137-001 and its whole stripped tails. The 
wavelength coverage is 4750 - 9350 \AA\ with a spectral resolution of R$\sim2600$ at the wavelength of H$\alpha$ emission 
line. As discussed in \citet{Boselli2021}, for the spectral resolution of \textit{MUSE}, the uncertainty of velocity dispersion 
is large when it is lower than $\sim$ 25-30 km s$^{-1}$. Since our studies focus on the diffuse ionized gas in stripped tails which 
typically has velocity dispersion larger than 30 km s$^{-1}$, the above limit should not influence the results and conclusions 
in this paper.

\footnotetext{https://esahubble.org/images/heic1404a/}
\begin{table*}
	\begin{center}
		\caption{The observational blocks used for making the mosaic of ESO~137-001 and its stripped tails}
		\label{tab:observational info.}
		\begin{tabular}{cccccccc}
			\hline
			OB & RA & Dec & Obs date & T$_{\rm exp}$ & Airmass & Seeing & Program ID / PI \\
			& (hh mm ss) & ($\degr$ $\arcmin$ $\arcsec$) & & (s) & & (arcsec) & \\
			\hline
			OB1    & 16 13 25.91 & -60 44 31.3 & February 6, 2020 & 720$\times$2  & 1.532-1.583 & 1.87-1.94 & 0104.A-0226(A) / Sun \\
			OB2-P1 & 16 13 24.70 & -60 45 39.1 & June 21, 2014 & 900$\times$3  & 1.239-1.251 & 0.83-0.96 & 60.A-9349(A) / SV\\
			OB2-P2 & 16 13 17.69 & -60 44 48.9 & June 21, 2014 & 900           & 1.238       & 0.78 & 60.A-9349(A) / SV    \\
			OB3    & 16 13 24.25 & -60 45 45.8 & June 18, 2017 & 900$\times$3  & 1.238-1.251 & 0.80-1.21 & 60.A-9100(G) / SV \\
			OB4    & 16 13 18.45 & -60 43 49.0 & February 4, 2020 & 720$\times$2  & 1.540-1.593 & 1.11-1.25 & 0104.A-0226(A) / Sun\\
			OB5    & 16 13 16.54 & -60 45 44.3 & May 20, 2015 & 1050$\times$2 & 1.238-1.243 & 1.22-1.35 & 095.A-0512(A) / Sun \\
			OB6    & 16 13 10.10 & -60 44 07.9 & May 21, 2015 & 1049$\times$2 & 1.238-1.241 & 1.52-1.88 & 095.A-0512(A) / Sun \\
			OB7-P1 & 16 13 10.09 & -60 45 06.9 & May 21, 2015 & 900$\times$2  & 1.347-1.380 & 0.81-1.00 & 095.A-0512(A) / Sun \\
			OB7-P2 & 16 13 02.12 & -60 44 20.8 & May 21, 2015 & 627$\times$2  & 1.418-1.451 & 1.04-1.12 & 095.A-0512(A) / Sun \\
			OB8    & 16 13 04.45 & -60 43 15.6 & February 5, 2020 & 720$\times$2  & 1.549-1.602 & 0.96-0.97 & 0104.A-0226(A) / Sun\\
			OB9    & 16 12 54.88 & -60 44 01.9 & March 17, 2020 & 720$\times$2  & 1.238-1.240 & 0.57-0.71 & 0104.A-0226(A) / Sun\\
			\hline
		\end{tabular}
	\end{center}
	\vspace{-0.3cm}
    \begin{tablenotes}
        \item Note: The observational blocks are ordered in right ascension from the East to the West. 
        The OB2 and OB7 both include two pointings. The previous works on ESO~137-001 \citep{Fumagalli2014,Fossati2016} 
        only used the data from 60.A-9349(A) to cover the galaxy and the front part of the primary tail. SV stands for 
        Science Verification.
    \end{tablenotes}
\end{table*}

For each pointing, we used the \textit{MUSE} pipeline (version 2.8.1; \citealt{Weilbacher2012,Weilbacher2020}) with the 
ESO Recipe Execution Tool (EsoRex; \citealt{ECDT2015}) to reduce the raw data, which provides a standard procedure to 
calibrate the individual exposures and combine them into a datacube (see the \textit{MUSE} pipeline 
manual\footnote{https://www.eso.org/sci/software/pipelines/muse/} for more details). Further sky subtraction 
was performed with the Zurich Atmosphere Purge software (ZAP; \citealt{Soto2016}), which involved principal component 
analysis to characterise the sky residuals remaining in the datacube and remove them. By using the CubeMosaic class 
implemented in the \textit{MUSE} Python Data Analysis Framework (MPDAF) package \citep{Bacon2016}, we combined the 
individual datacubes into the final mosaic. In this process, the individual datacubes were first manually aligned by 
adopting the bright \textit{2MASS} stars as a reference. Then, the $5\sigma$-clipping average was performed on them 
to build the final datacube mosaic and remove the outliers. We also propagated the variance and write them into the 
final datacube mosaic.

Before the spectral analysis of the final datacube mosaic, we corrected the foreground Galactic extinction of 
data based on the colour excess from the dust map of \citet{Schlegel1998} with the recalibration of \citet{Schlafly2011}. 
The Galaxy extinction law from \citet{Fitzpatrick1999} with $R_{V}=3.1$ is adopted in the above correction. 
Given the faintness of the diffuse emission and the seeing value, a median filter with a kernel of $4\times4$ spaxels 
(or 0$\farcs8\times0\farcs8$) was also chosen to smooth the final datacube mosaic. Then we adopted the public IDL 
software Kubeviz\footnote{https://www.mpe.mpg.de/~dwilman/kubeviz/Welcome.html} \citep{Fossati2016} to fit each emission 
line with the Gaussian profile and derived the emission-line fluxes, the velocity and velocity dispersion of the ionized 
gas. The spaxels with signal-to-noise ratio (S/N) $<3$ and velocity error and velocity dispersion error $>$ 50 km s$^{-1}$ 
were masked. In addition, we also masked the spaxels with extreme outliers in the distributions of velocity and velocity 
dispersion.

To measure the stellar kinematic properties in the galaxy disc and compare it with that of the ionized gas, we made 
a subcube to cover the D$_{25}$ (the isophotal level of 25 mag arcsec$^{-2}$ in the B band) of ESO~137-001.
The subcube has a spatial size of $100\arcsec\times50\arcsec$ and the same wavelength range as the above datacube mosaic. 
We computed the S/N of the 
stellar continuum within the wavelength range of 5300 - 5500 \AA\ and adopted the adaptive Voronoi method \citep{Cappellari2003} 
to spatially bin the data with a minimum S/N of 40. Then we use the software pPXF \citep{Cappellari2004,Cappellari2017} 
to fit the continuum and measure the stellar velocity and velocity dispersion for each bin. The continuum was modeled 
with the templates selected from the E-MILES simple stellar population 
models\footnote{http://research.iac.es/proyecto/miles/pages/spectral-energy-distributions-seds/e-miles.php} 
\citep{Vazdekis2016}, which are generated with a \citet{Chabrier2003} initial mass function and BaSTI isochrones 
\citep{Pietrinferni2004}. A total of 32 stellar templates were adopted, with eight ages increasing from 0.15 to 14 Gyr 
and four metallicities [M/H] ranging from -1.49 to 0.40. To avoid the sky residuals in the red part of \textit{MUSE} 
spectra, we limited the wavelength range within 4800 - 7000 \AA\ in the fitting process. We also extracted the integrated 
spectra within an aperture of $3\arcsec$ diameter and measured the system velocity 
of ESO~137-001 based on the stellar kinematics. The obtained system velocity is $4647\pm5$ km s$^{-1}$, which is adopted 
as the velocity reference of the stars and ionized gas throughout the paper.

\begin{figure*}
	\centering
	\includegraphics[width=\linewidth]{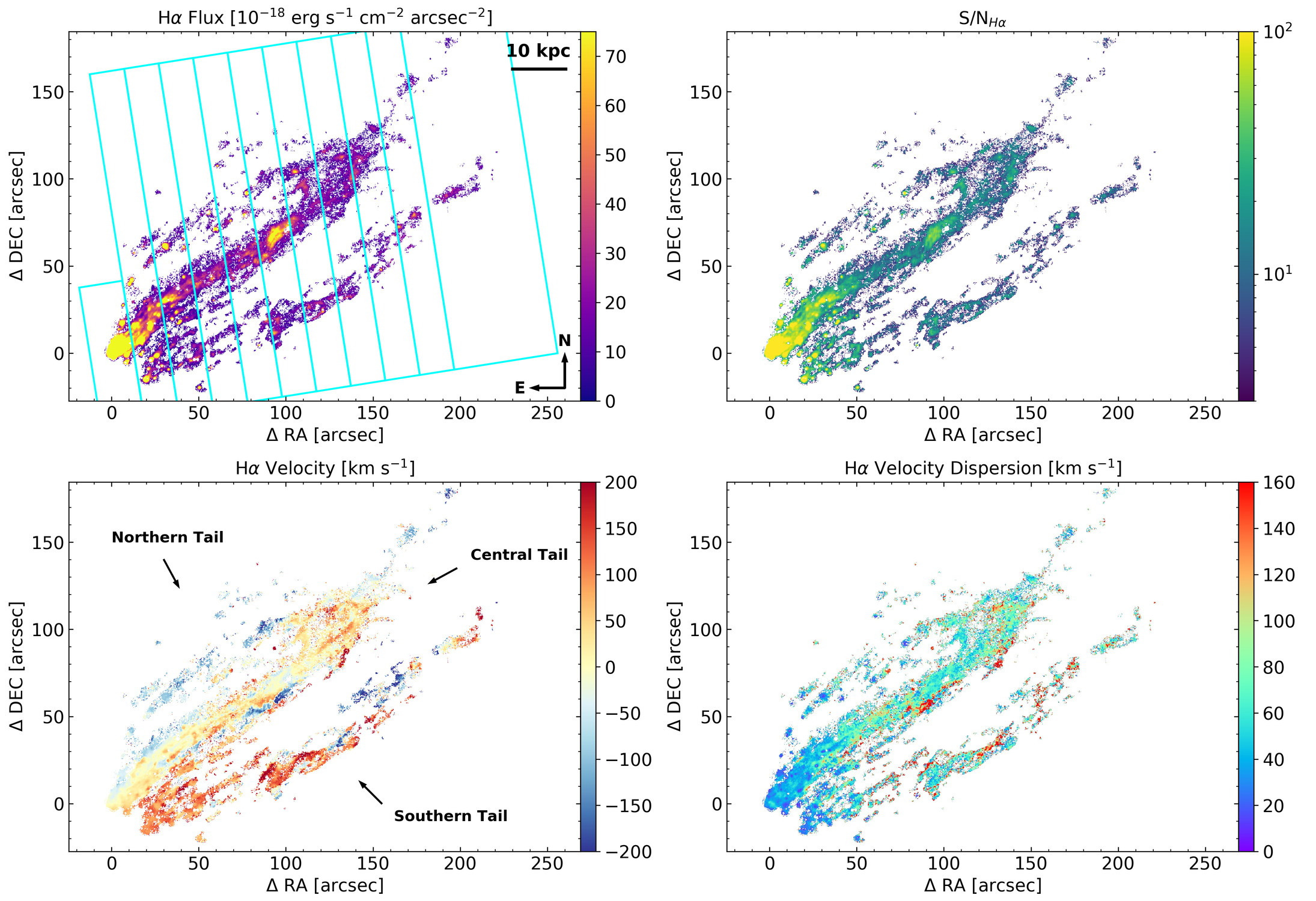}
        \vspace{-0.6cm}
	\caption{Two-dimensional maps for the properties of the ionized gas in ESO~137-001 and its stripped tails, 
	including the H$\alpha$ surface brightness and S/N, the velocity, and the velocity dispersion.
        The cyan boxes show the eleven regions where we performed the kernel density estimation. The major axis 
        of each region is paralleled with the major axis of the galaxy. The central tail, the southern tail, 
        and the northern tail are marked in the gas velocity field.}
	\label{fig:2d_maps_starPA}
\end{figure*}

\section{Results}
\label{sec:results}
\subsection{Overall distribution and kinematics of the ionized gas}
\label{subsec:overall properties}
In Fig.~\ref{fig:2d_maps_starPA}, we present the two-dimensional maps of the H$\alpha$ surface brightness (corrected for the 
Galactic extinction) and S/N, the velocity, and the velocity dispersion of the ionized gas in ESO~137-001 and its stripped tails. 
Our \textit{MUSE} observations provide a limit of $1.6\times10^{-18}$ erg s$^{-1}$ cm$^{-2}$ arcsec$^{-2}$ for the H$\alpha$ 
surface brightness at S/N $>3$, enabling us to trace the detailed distribution and structures of the ionized gas across the 
whole stripped tails. Based on the H$\alpha$ surface brightness, we separated \hii{} regions and the diffuse ionized 
gas\footnote{In the stripped tails, the ionized gas outside \hii{} regions is considered as the diffuse ionized gas. This 
could be different from the original concept of the diffuse ionized gas (DIG), which represents the diffuse ionized interstellar 
medium in the galaxy discs or haloes. The diffuse ionized gas in the stripped tails also includes the mixing component of the 
stripped ISM and the surrounding ICM.} in the stripped tails. The details for the identification of \hii{} regions were presented 
in \citet{Waldron2022}. In brief, we first adopted SExtractor with criteria on the shape of regions to identify the \hii{} region 
candidates. Then we applied the limits on the integrated H$\alpha$ flux 
and {\mbox{[N\,\textsc{ii}]}}/H$\alpha$ emission-line ratio to select the final \hii{} 
regions. In total, we selected 43 \hii{} regions in the stripped tails. They have a median H$\alpha$ surface brightness 
of $9.4\times10^{-17}$ erg s$^{-1}$ cm$^{-2}$ arcsec$^{-2}$, which is significantly higher 
than that of $1.4\times10^{-17}$ erg s$^{-1}$ cm$^{-2}$ arcsec$^{-2}$ for the diffuse ionized gas. 
All \hii{} regions are located within the inner half of the stripped tails and have a median projected distance 
of $\sim11$ kpc from the galaxy centre, suggesting their distributions tend to be close to the galaxy disc. 
This signature has also been noticed in the previous studies with the narrow band imaging and IFS 
observations \citep{Sun2007,Fossati2016}. At the middle part of the stripped tails ($\sim40$ kpc projected distance 
from the galaxy centre), one region with diffuse ionized gas also presents bright H$\alpha$ emission. As shown in \citep{Sun2021}, 
this region also has the highest X-ray surface brightness in the tail. There are no significant kinematic differences between 
this bright blob and its surrounding diffuse gas, suggesting it is just a denser concentration of gas with respect to the 
surrounding regions.

\begin{figure*}
	\centering
	\includegraphics[width=\linewidth]{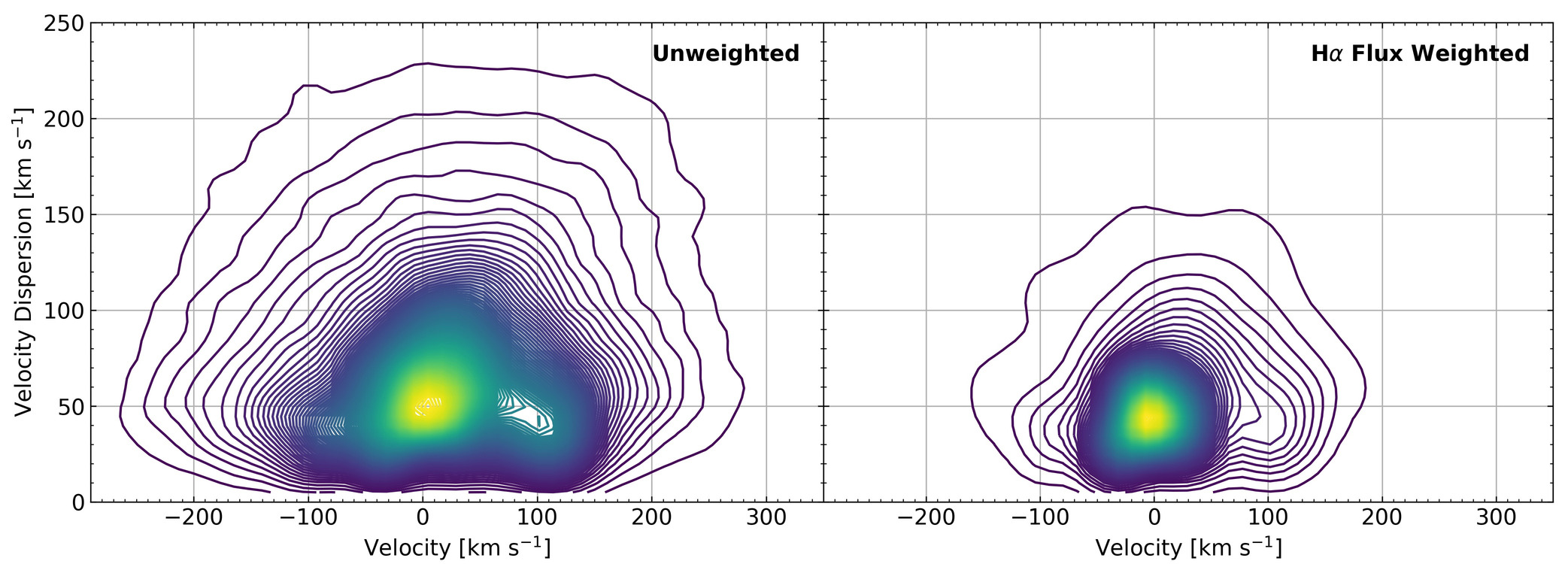}
        \vspace{-0.6cm}
	\caption{The kernel density estimation for the velocity vs. the velocity dispersion of the ionized gas 
	in ESO~137-001 and its stripped tails. The contours are separated by 1\%. The median velocity and velocity 
        dispersion are 17.3 km s$^{-1}$ and 62.5 km s$^{-1}$ respectively for unweighted, and 0.4 km s$^{-1}$ and 
        48.3 km s$^{-1}$ respectively for weighted.}
	\label{fig:kde_vel_dispersion_starPA}
\end{figure*}

\begin{figure*}
	\centering
	\includegraphics[width=\linewidth]{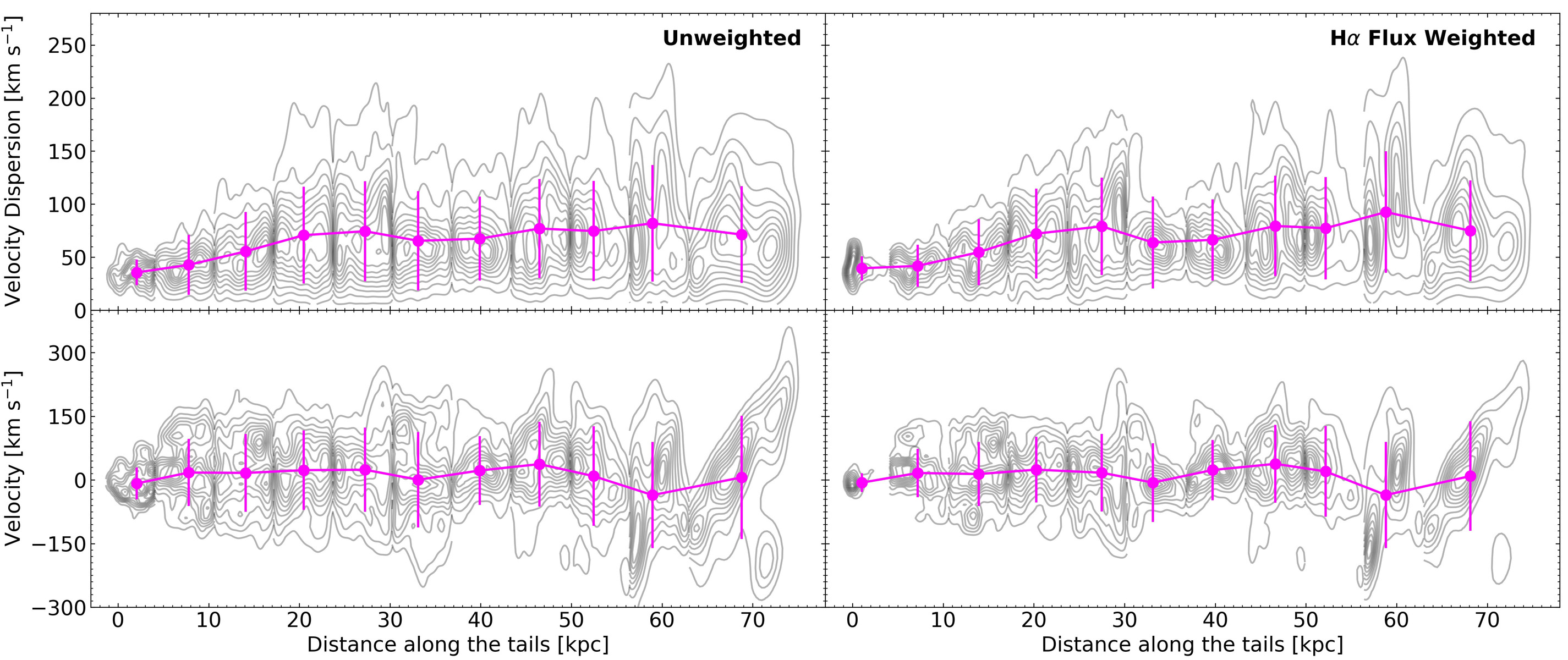}
        \vspace{-0.6cm}
	\caption{The kernel density estimation for the velocity and the velocity dispersion of the ionized gas 
	vs. the distance along the stripped tails. The contours are separated by 10\%. The magenta dots 
	show the median values $\pm$ 1 sigma scatter of the velocity and velocity dispersion in each region. 
        The median velocities are -8.3, 17.4, 16.5, 23.0, 24.0, 0.4, 22.2, 37.2, 9.0, -35.8 and 6.1 km s$^{-1}$ 
        (in the order of distance, from small to large) for unweighted, 
        and -6.4, 16.4, 13.9, 24.2, 16.9, -6.3, 23.4, 37.9, 20.0, -35.6 and 9.0 km s$^{-1}$ for weighted, 
        which suggests that most of the galaxy motion is on the plane of the sky.}
	\label{fig:kde_regions_starPA}
\end{figure*}

\begin{figure*}
	\centering
	\includegraphics[width=\linewidth]{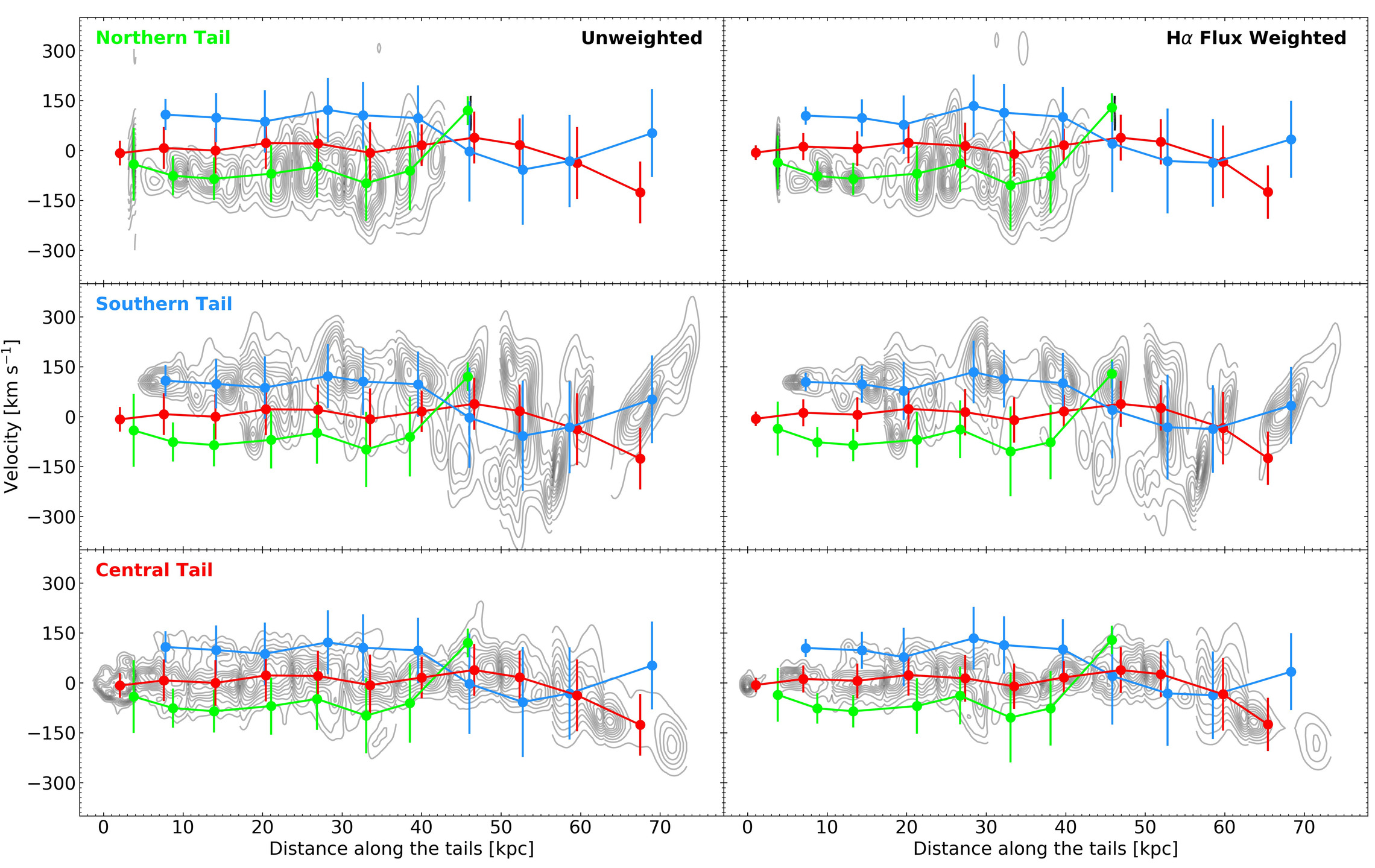}
        \vspace{-0.6cm}
	\caption{The kernel density estimation for the velocity of the ionized gas in three individual tails vs. 
        the distance along the stripped tails. The contours are separated by 10\%. The red, blue, and green dots 
        show the median values $\pm$ 1 sigma bars of the velocity in each region from the central, the southern, 
        and the northern tail, respectively. The three, initially kinematically distinct tails end with similar 
        velocities at distance $>$ 45 kpc.}
	\label{fig:kde_tails_vel_starPA}
\end{figure*}

\begin{figure*}
	\centering
	\includegraphics[width=\linewidth]{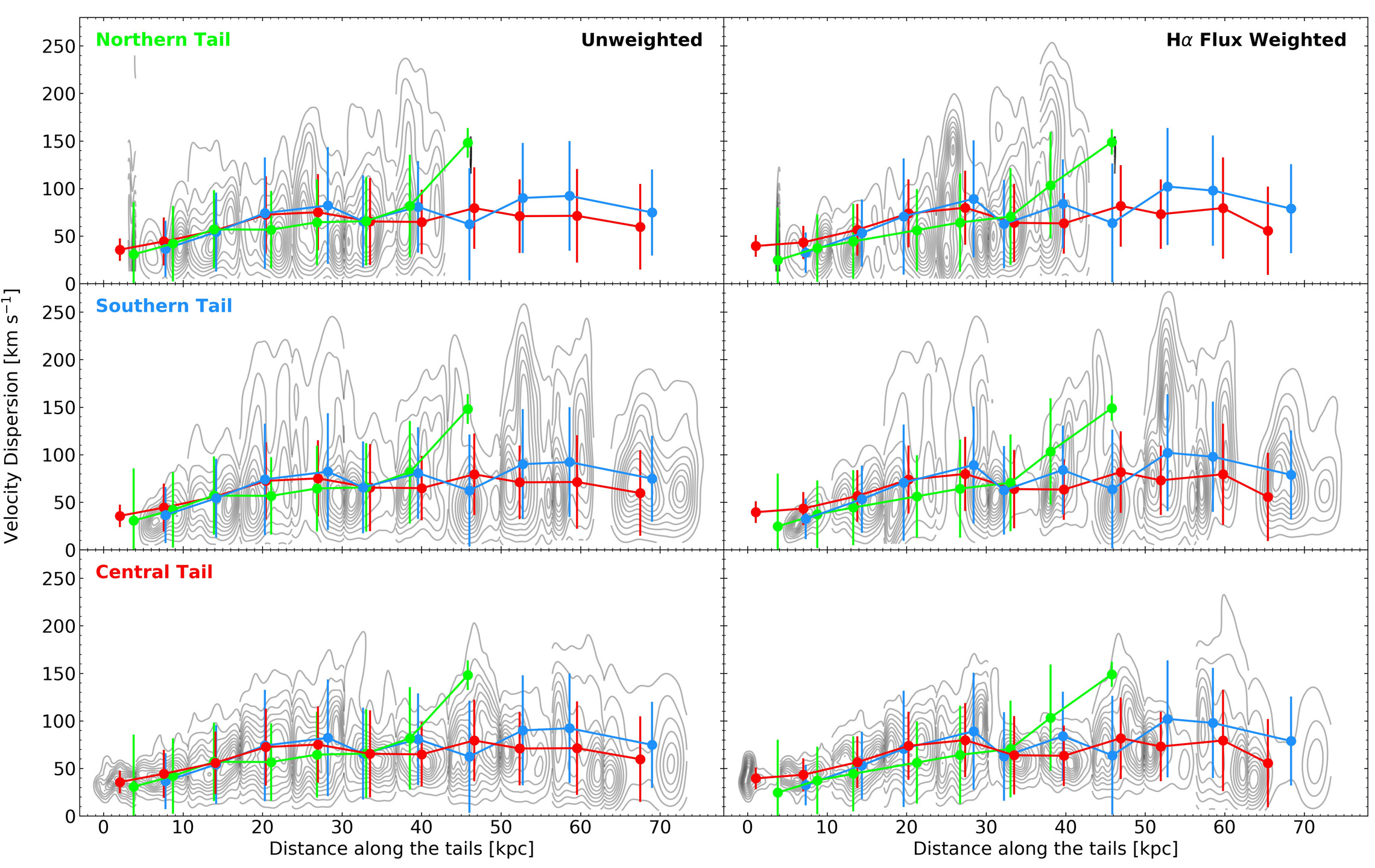}
	\vspace{-0.6cm}
	\caption{The kernel density estimation for the velocity dispersion of the ionized gas in each tail 
	vs. the distance along the stripped tails. The contours are separated by 10\%. The red, blue, 
	and green dots show the median values $\pm$ 1 sigma bars of the velocity dispersion in each region 
	from the central, the southern, and the northern tail, respectively. 
	}
	\label{fig:kde_tails_dispersion_starPA}
\end{figure*}

The stripped tails of ESO~137-001 are extended along the direction with a position angle (PA) 
of $\sim-54\degr$ (from the North, ``-'' for clockwise) and split 
into three main structures, including the central tail, the southern tail, and the northern tail. The central tail is 
the most extended and spread structure with a good continuity, which has an average width of $\sim10$ kpc and reaches to 
$\sim87$ kpc projected distance from the galaxy centre (16:13:27.231, -60:45:50.60) (The centre position was determined 
from the peaks of the X-ray, H$\alpha$ and CO emission in ESO~137-001, as well as 
the GALFIT fit to the \textit{HST} images, see \citealt{Waldron2022}). 
The southern tail also extends to $\sim75$ kpc projected distance from the galaxy centre, while its average width is 
only $\sim5$ kpc and the tail is not as continuous as the central tail. On the contrary, the northern tail is 
rather fragmented and presents several filamentary substructures. This tail covers a region of $\sim5$ kpc $\times$ 40 kpc.
The central tail is still connected with the inner galaxy disc while the northern and southern tails are not connected with 
it, suggesting they are probably at different stripping stages.

The velocity field of ionized gas presents a velocity gradient through the inner half of the stripped tails in ESO~137-001. 
From the northern tail to the southern tail, the LOS velocity of the ionized gas changes from blueshifted to redshifted and 
its median value ranges from $\sim-80$ km s$^{-1}$ to $\sim120$ km s$^{-1}$. This velocity pattern has also been observed 
by the spectroscopy of individual \hii{} regions \citep{Sun2010} and the previous IFS study \citep{Fumagalli2014}, which 
is considered as the rotation imprint of the galaxy disc. Beyond the inner half, the velocity field in the remaining 
part of the stripped tails becomes more disordered, indicating the kinematics state of the ionized gas could be different 
between these two regions. The velocity dispersion of ionized gas has a median value of $\sim30$ km s$^{-1}$ in the galaxy 
disc and \hii{} regions, while the diffuse ionized gas in the stripped tails presents an enhanced velocity dispersion 
to a median value of $\sim80$ km s$^{-1}$, suggesting it is probably more turbulent. 
Considering this kinematic difference between the \hii{} regions and diffuse ionized gas, we excluded \hii{} regions 
from the analysis in the rest of this section and Section \ref{subsec:velocity model}.
Note that several discrete regions with high velocity dispersion are elongated along the extension direction 
of the stripped tails and associated with kinematically distinct filamentary features in the velocity field. 
These may be distinct filamentary entities in the tails that are projected along the same LOS as other components, 
inducing the blending of emission lines with different velocities and yielding large values of velocity dispersion.
For the regions with velocity dispersion higher than 150 km s$^{-1}$, more than 80\% of them show the double-peak 
or non-Gaussian profiles for the emission lines. These regions only include less than 6\% spaxels of the whole 
datacube, which will not affect the following kinematic analysis.

We adopted the kernel density estimation to show the distributions of velocity and the velocity dispersion of the ionized 
gas in ESO~137-001 and its stripped tails. Kernel density estimation is a non-parametric way to estimate the probability 
density function, which allows us to visualize the underlying distribution of the variables in large datasets.
We present both the H$\alpha$ flux weighted and unweighted results. In general, the flux-weighted results show 
more concentrated distributions than the unweighted ones. Our conclusions are based on the flux-weighted results, while 
we also keep the unweighted results as a reference for the general distribution of the data, especially faint regions.
As shown in Fig.~\ref{fig:kde_vel_dispersion_starPA}, the velocity of most ionized gas ranges 
from $\sim-100$ km s$^{-1}$ to $\sim150$ km s$^{-1}$, while their velocity dispersion is mainly within 120 km s$^{-1}$.
In Fig.~\ref{fig:kde_regions_starPA}, we also present the kernel density estimation for the velocity and velocity 
dispersion of the ionized gas vs. the distance along the stripped tails. The estimation was performed 
in eleven individual regions with a direction parallel to the major axis of the galaxy. 
While we adopted $\sim8$ kpc as the width of the first region to cover the galaxy disc, the width of the regions 
along the stripped tails was set as $\sim6.5$ kpc. For the last region, we used $\sim20$ kpc as its width 
to collect enough data points. The exact positions of these regions are shown as cyan boxes in Fig.~\ref{fig:2d_maps_starPA}.
The median velocity of the ionized gas in the stripped tails does not significantly change with the distance, which is around 
15 km s$^{-1}$. This result confirms that the motion of ESO~137-001 is mainly on the sky plane, which has also been mentioned 
by several previous studies \citep{Sun2006,Sun2007,Sun2010,Fumagalli2014,Jachym2019}. In addition, the small redshifted median 
velocity suggests that ESO~137-001 is also slightly moving towards us along the LOS. This is consistent with the 
implication from the radial velocity of ESO~137-001, which is $\sim224$ km s$^{-1}$ smaller than the Norma cluster's velocity.
The velocity spread of ionized gas is $\sim350$ km s$^{-1}$ in the stripped tails, which is significantly larger than that 
of $\sim120$ km s$^{-1}$ in the galaxy disc. While this result is generally consistent with the behaviours predicted by 
the simulations (e.g, \citealt{Roediger2008}), the small fluctuations of velocity spread along the tails do not suggest 
strong turbulence there. The median velocity dispersion of ionized gas changes 
from $\sim35$ km s$^{-1}$ in the galaxy disc to $\sim80$ km s$^{-1}$ at $\sim20$ kpc downstream and maintain 
a similar level in the remaining part of the stripped tails. The spread of velocity dispersion also presents a similar 
behaviour, which increases from $\sim50$ km s$^{-1}$ in the galaxy disc to $\sim200$ km s$^{-1}$ beyond $\sim20$ kpc 
downstream of the stripped tails. These results suggest that the degree of turbulence in the diffuse ionized gas is 
probably enhanced from the galaxy disc to $\sim20$ kpc along the stripped tails.

To further explore the kinematic differences between the three split tails, we performed similar kernel 
density estimations as above for each of them. The results are shown in Fig.~\ref{fig:kde_tails_vel_starPA} 
and \ref{fig:kde_tails_dispersion_starPA}. 
Within $\sim$ 40 kpc, the three tails maintain distinctive velocities at $\sim-80$ km s$^{-1}$ to $\sim120$ km s$^{-1}$, 
which further clearly shows the impact of the galactic rotation on the front part of the tails. 
Beyond the distance of $\sim40$ kpc, the median velocity of ionized gas in different tails becomes about the same, 
indicating the disappearance of the velocity gradient. The median velocity dispersion of ionized gas in different tails 
presents a similar behaviour as that in Fig.~\ref{fig:kde_regions_starPA}, further supporting the enhanced degree of 
turbulence from the galaxy disc to $\sim20$ kpc along the stripped tails. In addition, the spreads of the velocity and 
velocity dispersion of the ionized gas are also comparable in different tails. These results suggest that the general 
kinematic states of the ionized gas in three split tails are probably similar, except for the pattern of the velocity 
gradient across them. Note that different calculations of the distance and the direction of regions do not significantly 
affect the results of the above kernel density estimations. We have performed these analyses based on different definitions 
of the distance (e.g., the distance to the galaxy centre/major axis) or different directions of the regions (e.g., the major 
axes of the regions perpendicular to the extension direction of the stripped tails). The general kinematic behaviours 
of the ionized gas are similar to the above one in each case. 

\subsection{The distributions and velocity fields of the ionized gas and stars in the galaxy disc}
\label{subsec:kinematics in the galaxy disc}
\begin{figure*}
	\centering
	\includegraphics[width=0.95\linewidth]{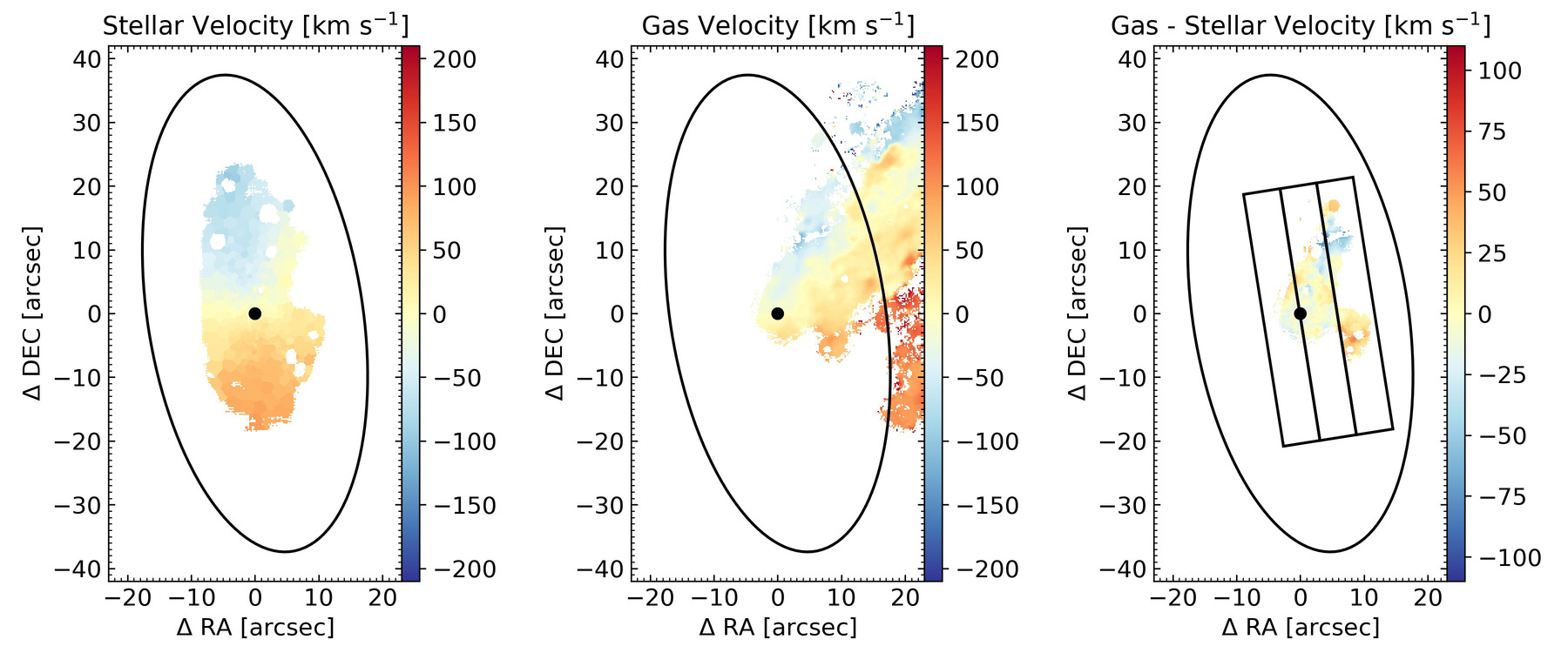}
	\vspace{-0.3cm}
	\caption{The comparison between the distributions and velocity fields of the ionized gas and stars in the galaxy 
        disc of ESO~137-001. The foreground stars are masked in the stellar velocity field. The position of the galaxy 
        centre is marked as the black circle. The black ellipses present the D$_{25}$ of the galaxy. The black boxes show 
        the three regions where we calculate the error-weighted averages of the velocity differences between the ionized 
        gas and stars. The major axes of these regions are paralleled with the major axis of the galaxy. When the gas is 
        decelerated by ram pressure, it is expected that gas should have a higher radial velocity than the galaxy, 
        as indeed observed here.}
	\label{fig:stellar_gas_vel_bin}
\end{figure*}

To understand the state of ionized gas in the galaxy disc of ESO~137-001, we compared its distribution and velocity field 
with those of the stars. As shown in Fig.~\ref{fig:stellar_gas_vel_bin}, the stars are well distributed in a disc shape 
and present a regular rotation motion. To measure the rotation curve of the stars, we adopted the tilted-ring method (e.g, \citealt{Rogstad1974,Bosma1978,Begeman1987,Schoenmakers1997,Wong2004,Krajnovic2006}) 
to fit the stellar velocity field. In this method, the projected galaxy disc is divided into a set of elliptical annuli. 
For each annulus, the LOS velocity is fitted as a Fourier series, which describes the kinematic components with different 
azimuthal symmetries:
\begin{equation}
V_{\rm LOS}(R,\psi) = V_{\rm sys}+\sum\limits_{n=1}^k[c_n(R){\rm sin}(n\psi)+s_n(R){\rm cos}(n\psi)].
\end{equation}
($R$, $\psi$) represent the polar coordinate in the plane of the galaxy disc. V$_{\rm sys}$ is the system velocity of the 
galaxy. The first-order terms $c_1(R){\rm sin}\psi$ and $s_1(R){\rm cos}\psi$ describe the radial and rotational velocity 
components, while the higher-order terms could be related to the perturbations of gravitational potential 
(e.g., \citealt{Schoenmakers1997,Wong2004,Krajnovic2006}). Thus, the stellar rotation curve can be derived as: 
\begin{equation}
V_{\rm rot}(R) = s_1(R)/{\rm sin}i,
\end{equation}
where $i$ is the inclination of the galaxy.

\begin{figure}
	\centering
	\includegraphics[width=0.95\linewidth]{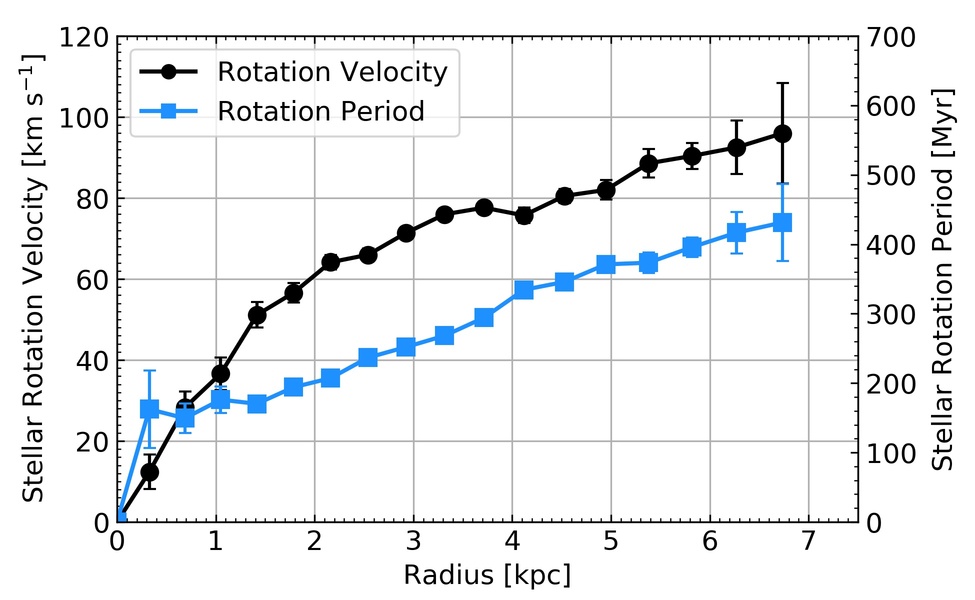}
	\vspace{-0.3cm}
	\caption{The stellar rotation curve of ESO~137-001 and the corresponding stellar rotation periods 
        as a function of the radius.}
	\label{fig:stellar_rotation_curve}
\end{figure}

We used the software package Kinemetry \citep{Krajnovic2006} to fit the LOS velocity with a three-order Fourier series. 
The PAs and inclinations of the elliptical annuli are fixed at $9\degr$ and $66\degr$, which are obtained with the 
photometric analysis of the \textit{HST} images \citep{Waldron2022}. The galaxy centre was fixed at 
(16:13:27.231, -60:45:50.60) (see Section \ref{subsec:overall properties}) and we adopted 
the system velocity derived in Section \ref{sec:observations and data processing}. The radial distribution of 
s$_1$ coefficient was used to derive the stellar rotation curve, which is shown in Fig.~\ref{fig:stellar_rotation_curve}. 
Because of the limited S/N of the observed continuum spectra, we can only measure the stellar rotation velocity up to the 
radius of $\sim6.7$ kpc. Based on the rotation curve, we also calculate the corresponding stellar rotation periods as a 
function of the radius (Fig.~\ref{fig:stellar_rotation_curve}).

\begin{figure}
	\centering
	\includegraphics[width=0.95\linewidth]{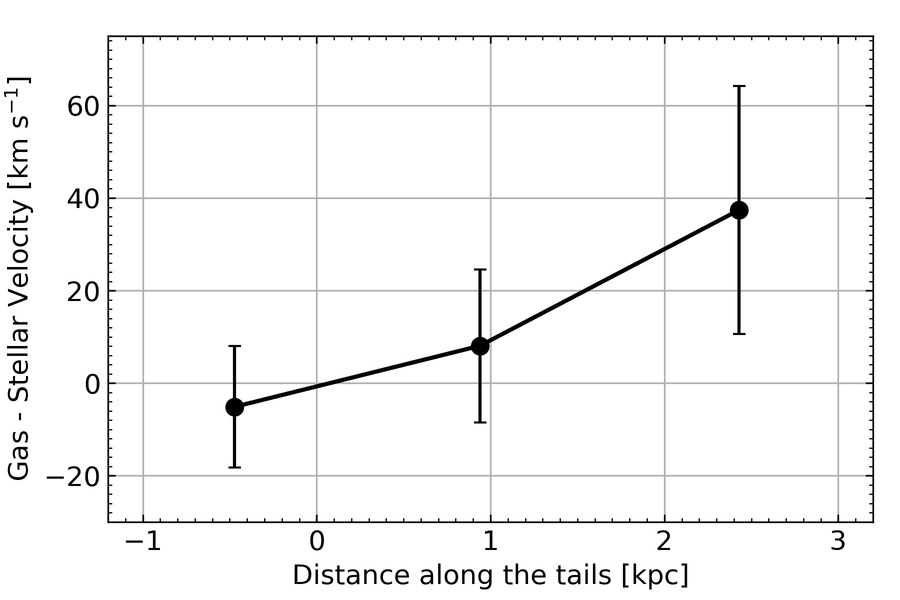}
	\vspace{-0.3cm}
	\caption{The error-weighted averages of the velocity differences between the ionized gas 
        and stars as a function of the distance along the stripped tails. These averages are calculated 
        in three regions across the galaxy disc, which are shown with black boxes in the right panel of 
	    Fig.~\ref{fig:stellar_gas_vel_bin}.}
	\label{fig:stellar_gas_waverage}
\end{figure}

In Fig.~\ref{fig:stellar_gas_vel_bin}, we present the velocity fields of the ionized gas and stars in the galaxy 
disc and make a comparison between them. The overall size of the ionized gas in the galaxy is drastically decreased, 
suggesting most gas component in the outer galaxy disc has been stripped. Ram pressure shapes the remnant ionized 
gas into a narrow cone, which connects the inner galaxy disc with the central tail. This connection indicates that 
the ionized gas in the inner galaxy disc could be the origin of the central tail and still feed the stripping 
process there. To clearly present the kinematic difference, we further calculate the error-weighted averages of the 
velocity differences between the ionized gas and stars in three regions across the galaxy disc. As shown in 
Fig.~\ref{fig:stellar_gas_waverage}, there is an increasing of the velocity difference from $\sim0$ km s$^{-1}$ 
in the inner galaxy disc to $\sim40$ km s$^{-1}$ in the outer galaxy disc. Considering the galaxy has a small velocity 
component towards us, this redshifted velocity difference provides a potential hint of the gas deceleration due to the 
RPS. The above significant difference between the ionized gas and stars suggests the intense perturbations produced by 
the RPS and this process is already well advanced. 

Numerical simulations have predicted the influence of RPS on the gas kinematics of galaxies, which includes changing 
the rotation curves, producing the mismatch between the kinematic centres of stars and gas components, unwinding the 
spiral arms, and inducing distortions in the velocity fields (e.g., \citealt{Kronberger2008,Haan2014,Bellhouse2021}). 
Regarding observational studies, spatially-resolved spectroscopy can provide the detailed kinematic properties 
of RPS galaxies, which have shown several examples of the above effects. For the RPS galaxy SOS~114372, 
\citet{Merluzzi2013} observed the one-sided extraplanar ionized gas presents the rotation motion up to a projected 
distance of $\sim13$ kpc from the galaxy disc, while the distribution and kinematics of the stars are symmetric and 
regular. Based on the \textit{MUSE} and Fabry-Perot observations of galaxy IC 3476, \citet{Boselli2021} discovered 
the kinematic centre of the ionized gas has an offset of $\sim500$ pc from that of the stars, due to the perturbations 
induced by the RPS process. In addition, they also found that the mean velocity of the gas with respect to that of 
the stars has been changed by $\sim20$ km s$^{-1}$. As shown in the GASP survey, the perturbations of velocity fields 
have also been observed in the ionized gas of several RPS galaxies, for example, J201 and J204 
(e.g., \citealt{Bellhouse2017,Gullieuszik2017}). 

The comparison between ESO~137-001 and the above targets indicates more intense effects of RPS on the gas kinematics 
of this galaxy, since the gas in the outer disc of ESO~137-001 has been fully removed while the loss of the gas content 
in other galaxies is relatively mild. The influence of RPS on gas kinematics depends on several factors, including 
the mass of the gas content, the gravitational potential of the RPS galaxy, the strength of the ram pressure, 
the geometrical configuration, and the stage of the stripping process. Both face-on and edge-on stripping processes exist 
in these galaxies, suggesting the geometrical configuration is not a key factor. In addition, the gas and stellar mass 
of ESO~137-001 are not outstanding among these galaxies. We thus propose the strength of the ram pressure and the stage 
of the stripping process may play important roles in shaping the intense RPS effects of ESO~137-001.

\begin{figure*}
	\centering
	\includegraphics[width=\linewidth]{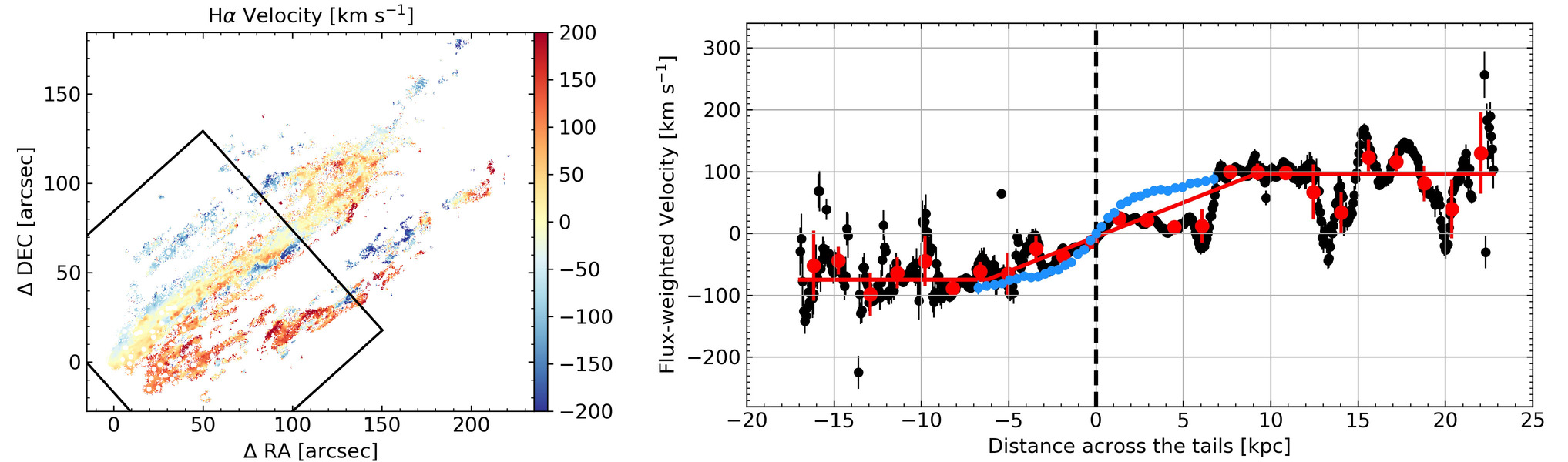}
	\vspace{-0.6cm}
	\caption{Fitting the global velocity gradient of the ionized gas in the stripped tails of ESO~137-001. 
        Left panel: The velocity field of the ionized gas. The selected region with a clear velocity gradient 
        is marked by the black box. Right panel: The black dots show the averaged velocities, weighted by the 
        H$\alpha$ flux from spaxels in each column, while their error-weighted averages in twenty-five distance 
        bins are presented in red dots. The red solid line describes the best-fitting global velocity gradient. 
        The blue dots present the stellar rotation curve of the galaxy (without the projection correction). 
        The position of the galaxy centre is indicated with the black dashed line.}
	\label{fig:gasvel_modelling_process}
\end{figure*}

\begin{figure*}
	\centering
	\includegraphics[width=\linewidth]{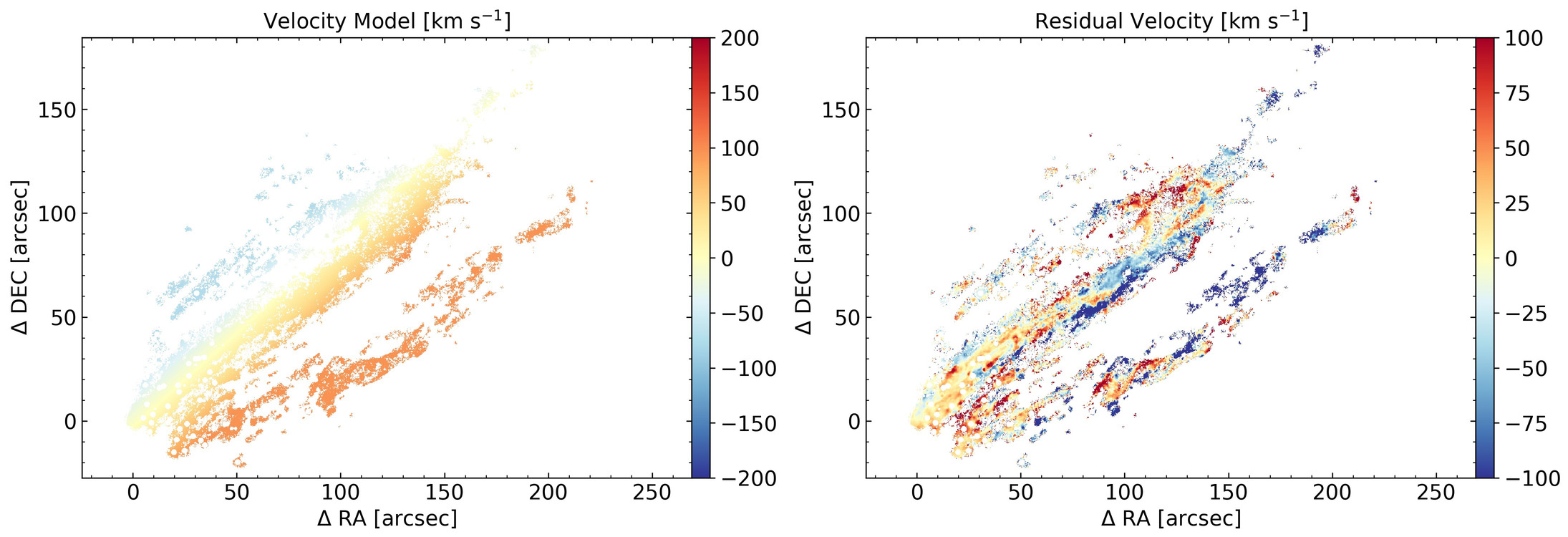}
	\vspace{-0.6cm}
    \caption{The velocity model of the ionized gas in ESO~137-001 and its stripped tails and the corresponding residual 
	    velocity field. The residual velocity field is obtained by subtracting the velocity model from the observed velocity 
	    field of the ionized gas.}
	\label{fig:gasvel_modelling_results}
\end{figure*}

\begin{figure*}
	\centering
	\includegraphics[width=\linewidth]{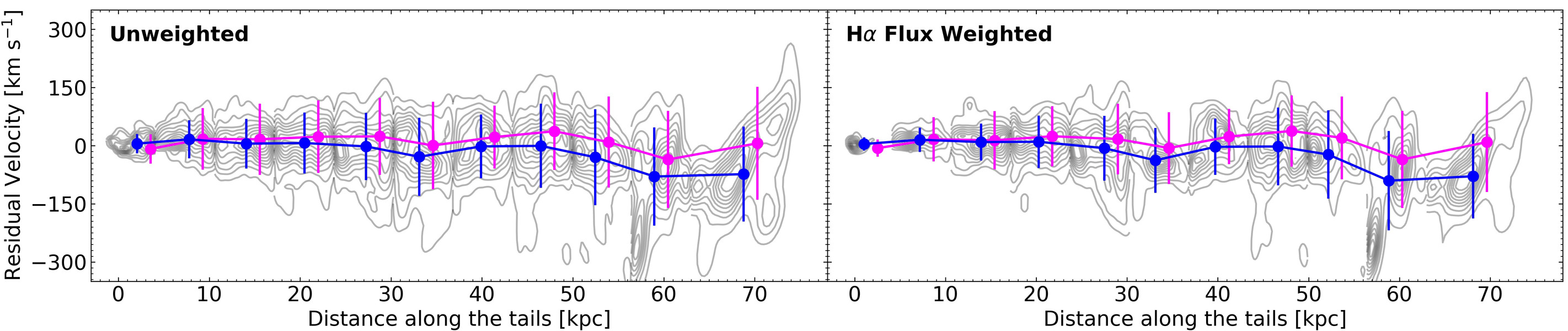}
	\vspace{-0.6cm}
    \caption{The kernel density estimation for the residual velocity (the right panel of 
        Fig.~\ref{fig:gasvel_modelling_results}) vs. the distance along the stripped tails. 
	    The contours are separated by 10\%. The blue dots show the median values $\pm$ 1 sigma bars of the residual 
	    velocity in each region. The RMS of the median residual velocity is 35.5 km s$^{-1}$ for unweighted and 
        39.1 km s$^{-1}$ for weighted. The magenta dots present the median velocities $\pm$ 1 sigma bars as shown 
	    in Fig.~\ref{fig:kde_regions_starPA}, which are offset for comparison. The medians of the scatter ratios 
	    between the two curves (blue/magenta) are 0.87 for unweighted and 0.9 for weighted.}
	\label{fig:reskde_regions_vel}
\end{figure*}

\begin{figure*}
	\centering
	\includegraphics[width=\linewidth]{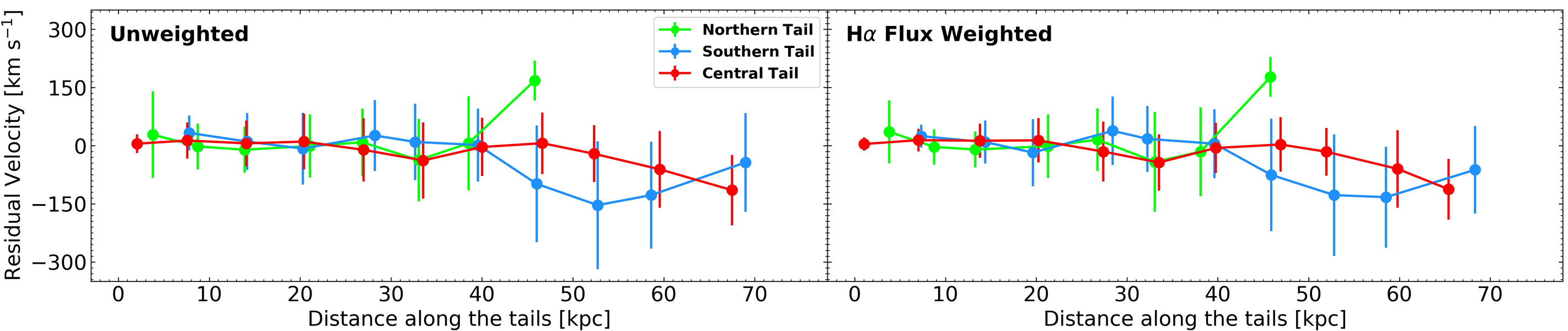}
    \vspace{-0.6cm}
	\caption{The median value of residual velocity in each tail vs. the distance along the stripped tails. 
	    The red, blue, and green dots show the median values $\pm$ 1 sigma bars of the residual velocity in each 
        region from the central, the southern, and the northern tail, respectively. While the rotation pattern 
        in the front part of the tails (within 40 kpc from the galaxy) is well removed, it over-corrects the regions beyond.}
	\label{fig:reskde_tails_vel}
\end{figure*}

\subsection{Modelling the velocity field of the ionized gas}
\label{subsec:velocity model}
To further quantify the kinematic state of the stripped tails in ESO~137-001 and the development of turbulence in the 
stripped gas, the velocity gradient from the galactic rotation needs to be first modeled and removed. As the galactic 
rotation imprint began to disappear as a function of the distance to the galaxy, the velocity field of the stripped 
gas is more at the mercy of turbulence in the wake. We constructed the velocity model for the residual galactic 
rotation based on the velocity gradient of the ionized gas. Since the kinematic properties of the ionized gas 
are not uniform along the stripped tails, we attempted two different methods:
(1) select the front part of the tails with a clear velocity gradient and use it as the global velocity gradient; 
(2) divide the stripped tails into some regions along the tails and measure the velocity gradient in each region separately.

\subsubsection{Modelling based on the global velocity gradient}
\label{subsubsec:velocity model:method one}
As shown in Fig.~\ref{fig:gasvel_modelling_process}, to derive the global velocity gradient, we selected the region 
including the head and inner half of ESO~137-001's stripped tails (within $\sim40$ kpc from the major axis of the galaxy disc). 
To determine the best-fitting direction and slope of the global velocity gradient, we rotated the velocity field within the 
selected region at different angles and correspondingly calculated the H$\alpha$ flux-weighted averages of the velocities from 
the spaxels in each column. For each rotation angle, the velocity gradient was obtained by fitting the above average velocities 
as a function of the distance across the stripped tails. We adopted a piecewise function including a linear function 
and two plateaus as the fitting function (see the right panel of Fig.~\ref{fig:gasvel_modelling_process}). 
Due to the statistical fluctuations in the data, the error-weighted averages in twenty-five distance bins 
were used in the fitting process. The best-fitting direction of the global velocity gradient was determined to have the lowest 
root mean square (RMS) in the corresponding residual velocity field (velocity field – velocity gradient), which is along the 
direction with a PA of $42\pm1\degr$. The best-fitting slope of the global velocity gradient is $11.1\pm1.1$ km s$^{-1}$ kpc$^{-1}$ 
between $-74.9$ km s$^{-1}$ and 96.0 km s$^{-1}$ at the distance of $-6.2$ kpc and 9.2 kpc. The velocity range of this 
gradient is generally consistent with that of the stellar rotation curve, which further confirms it is the rotation imprint 
of the galaxy. For the whole velocity field, we obtained the model by applying the above velocity gradient 
to the rest of the stripped tails. 

A possible caveat in the above analysis is that the fitting function might be too simple to well describe the gas velocity field. 
Our aim to fit the gas velocity field is to quantify the large-scale velocity gradient across the stripped tails. The choice 
of fitting function is based on the fact that this large-scale velocity gradient itself can be described as a linear function. 
On the other hand, there is no exact model for the large-scale velocity gradient across the stripped tails. In this case, 
the linear function could be a realistic description of a velocity gradient. On the small scale, the velocity fluctuations 
always exist in the turbulent environment of stripped tails. A more complicated function can exactly quantify these small-scale 
velocity fluctuations. However, this deviates from our research goal to quantify the large-scale velocity gradient.

Fig.~\ref{fig:gasvel_modelling_results} presents the velocity field model and its corresponding residual velocity field. 
While the rotation imprint is well removed in the front part of the stripped tails (or where the model is constructed), 
the residual velocity significantly increases in the rear half of the stripped tails. To quantitatively examine the velocity 
field model, we performed the kernel density estimation for the residual velocity vs. the distance along the stripped tails. 
The estimation was carried out in the same regions as defined in Section \ref{subsec:overall properties}. As shown in 
Fig.~\ref{fig:reskde_regions_vel} and \ref{fig:reskde_tails_vel}, the median values of the residual velocity are close to zero 
within the inner $\sim40$ kpc, indicating the global velocity gradient of the ionized gas has been well modeled and subtracted 
from the observed velocity field. There is still considerable residual velocity beyond the distance of $\sim40$ kpc, presumably 
because the residual rotation no longer dominates the kinematics of the ionized gas there, while other motions such as turbulence 
may begin to play a role.

The kinematics of stripped tails are closely related to the original kinematic state of the ISM in galaxy discs and 
the effects of RPS. Since ESO~137-001 is a late-type spiral galaxy, the kinematics of the ionized gas in the galaxy disc 
should generally follow the stellar kinematics there. While the direction of the global velocity gradient clearly deviates from 
the direction of the major axis of the stellar disc (PA: $42\degr$ vs. $9\degr$). This result highlights the influence of RPS on 
the kinematic properties of stripped gas. Considering this point, for ESO~137-001, we suggest it is not sufficient to adopt the 
stellar velocity field itself as a reference to model the velocity field of ionized gas in the stripped tails. Although the 
velocity gradient in the stripped tails could be originated from the rotation motion of the galaxy disc, the velocity vector 
in the direction where RPS acts must be taken into account.

\begin{figure*}
	\centering
	\includegraphics[width=\linewidth]{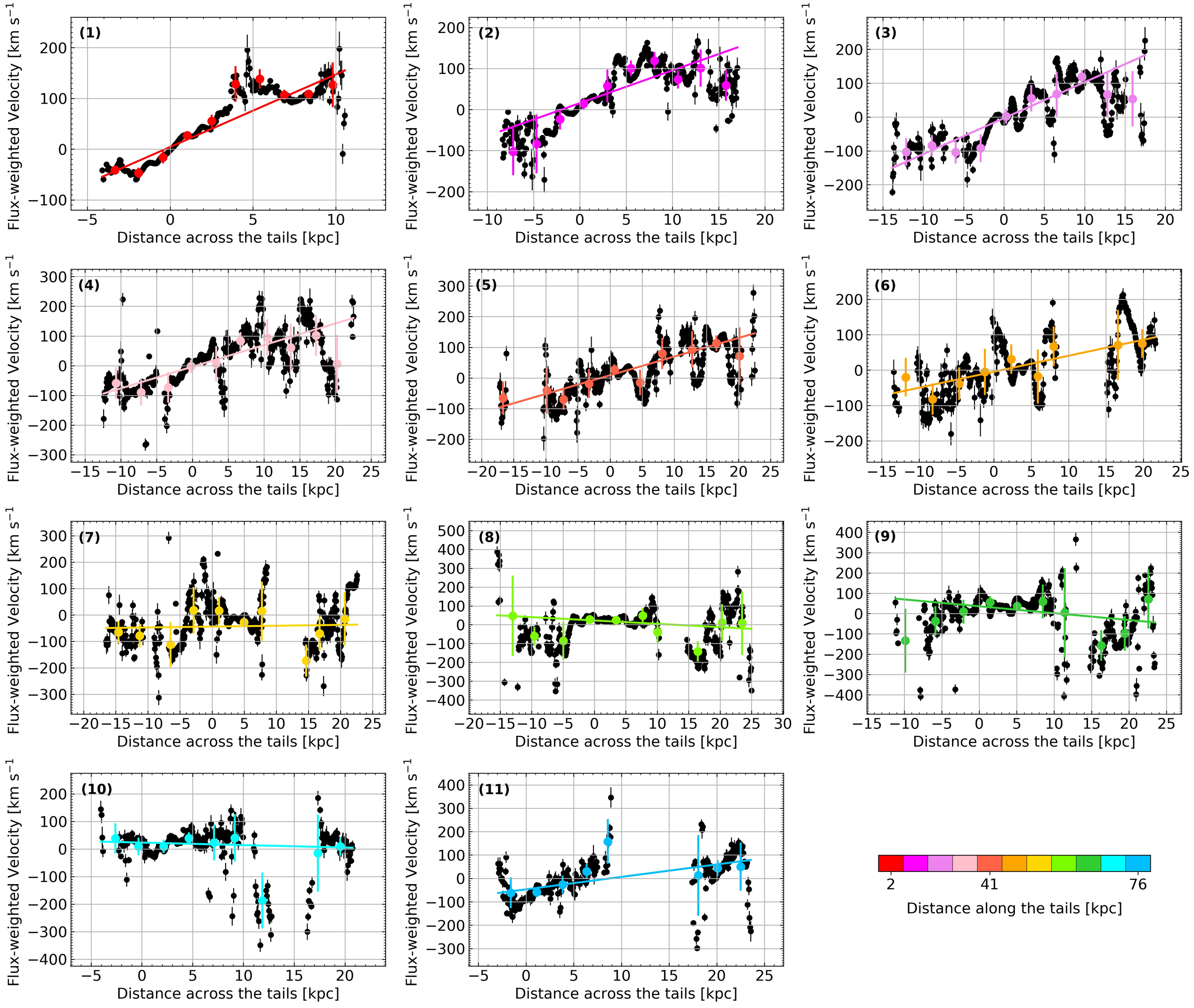}
    \vspace{-0.6cm}
	\caption{Fitting of the velocity gradient of the ionized gas in divided regions along the stripped tails of ESO~137-001. 
	The positions of these regions are indicated by different colours. For each region, the black dots present the H$\alpha$ 
	flux-weighted averages of the velocities from the spaxels in each column. The coloured dots show the error-weighted 
	averages of the data in ten distance bins. The coloured line describes the best-fitting velocity gradient.}
	\label{fig:rotation_curves_fitting}
\end{figure*}

\begin{figure*}
	\centering
	\includegraphics[width=\linewidth]{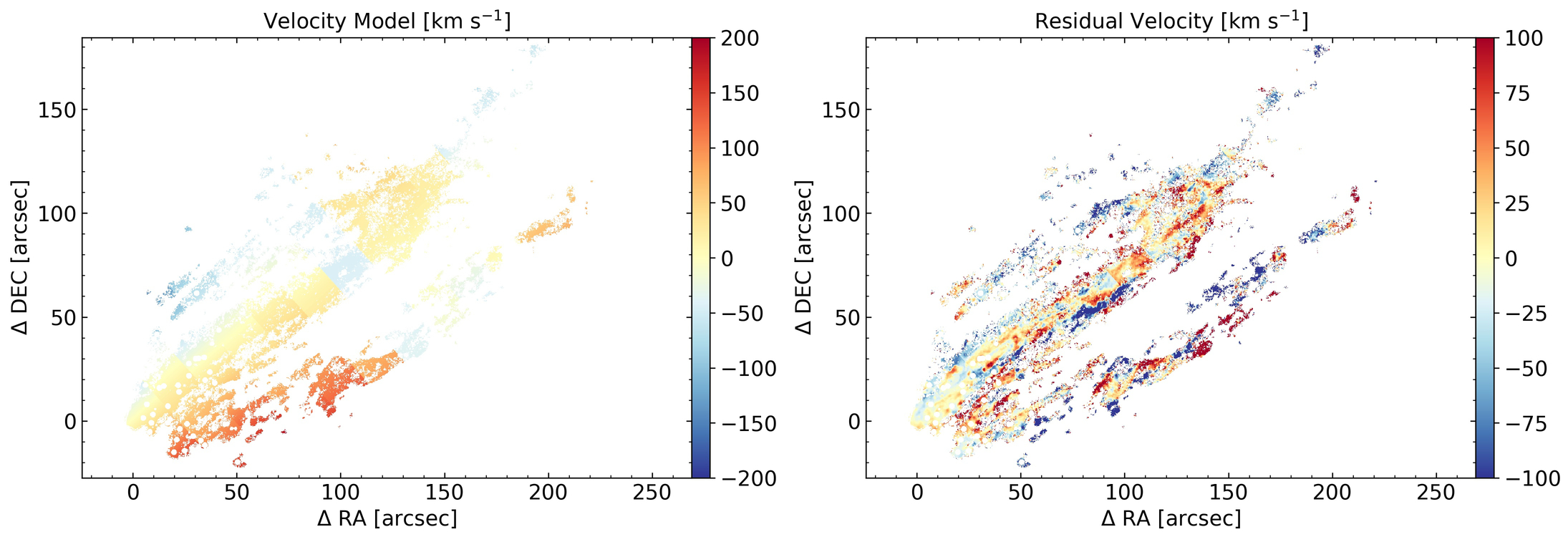}
	\vspace{-0.7cm}
	\caption{The velocity model of the ionized gas in ESO~137-001 and its stripped tails and the corresponding residual velocity field. 
	The residual velocity field is obtained by subtracting the velocity model from the observed velocity field of the ionized gas.}
	\label{fig:gasvel_modelling_results_velgradient}
\end{figure*}

\begin{figure*}
	\centering
	\includegraphics[width=\linewidth]{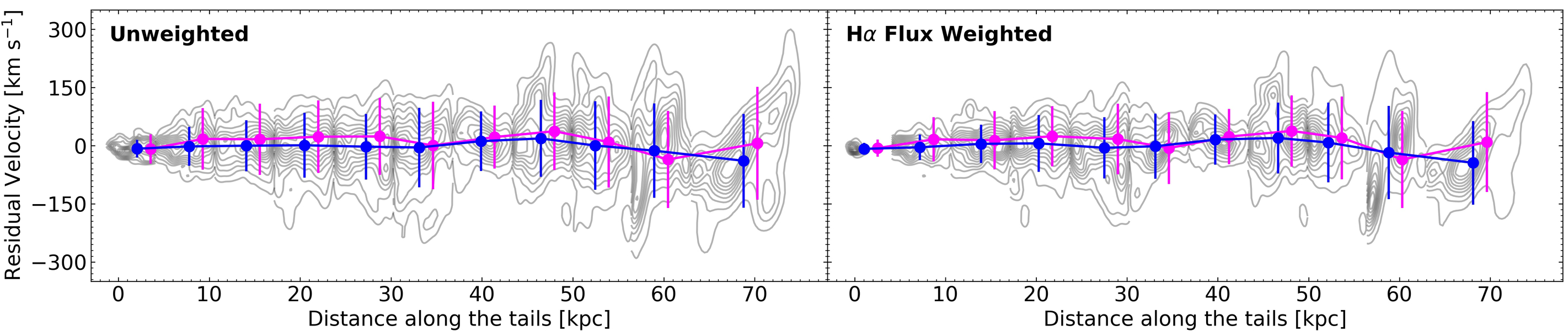}
	\vspace{-0.6cm}
	\caption{The kernel density estimation for the residual velocity (the right panel of 
        Fig.~\ref{fig:gasvel_modelling_results_velgradient}) vs. the distance along the stripped tails. 
	    The contours are separated by 10\%. The blue dots show the median values $\pm$ 1 sigma bars of the 
        residual velocity in each region. The RMS of the median residual velocity are 14.4 km s$^{-1}$ for 
        unweighted and 16.9 km s$^{-1}$ for weighted, which are significantly lower than those in the first 
        method shown in Fig.~\ref{fig:reskde_regions_vel} (as here blue points are close to zero than those 
        in Fig.~\ref{fig:reskde_regions_vel}). The magenta dots present the median velocities $\pm$ 1 sigma 
        bars as shown in Fig.~\ref{fig:kde_regions_starPA}, which are offset for comparison. The medians of 
        the scatter ratios between the two curves (blue/magenta) are 0.89 for unweighted and 0.91 for weighted.}
	\label{fig:reskde_regions_vel_method2}
\end{figure*}

\begin{figure*}
	\centering
	\includegraphics[width=\linewidth]{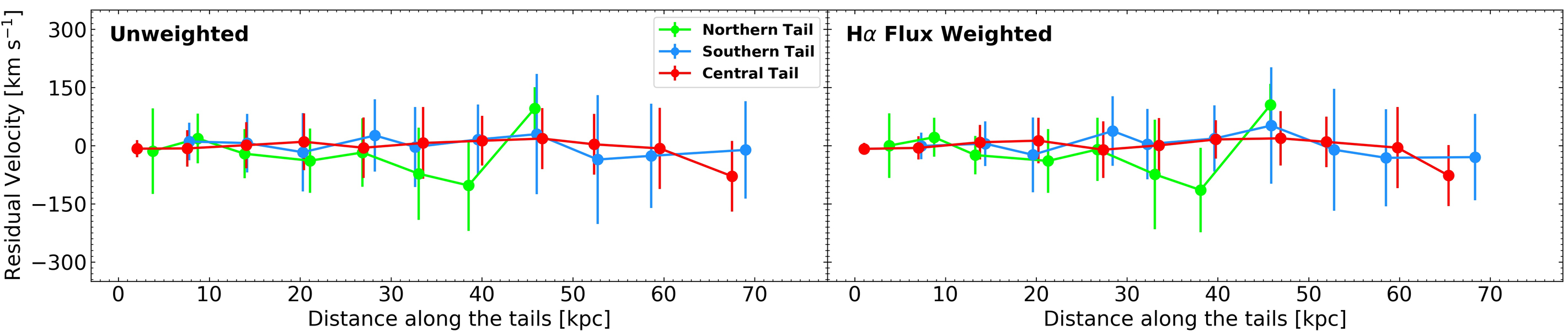}
	\vspace{-0.6cm}
	\caption{The median value of residual velocity in each tail vs. the distance along the stripped tails. 
        The red, blue, and green dots show the median values $\pm$ 1 sigma bars of the residual velocity in each 
        region from the central, the southern, and the northern tail, respectively. With the correction derived from 
        Fig.~\ref{fig:rotation_curves_fitting}, indeed the median velocities of different tails are now all consistent 
        with zero.}
	\label{fig:reskde_tails_vel_method2}
\end{figure*}

\begin{figure}
	\centering
	\includegraphics[width=\linewidth]{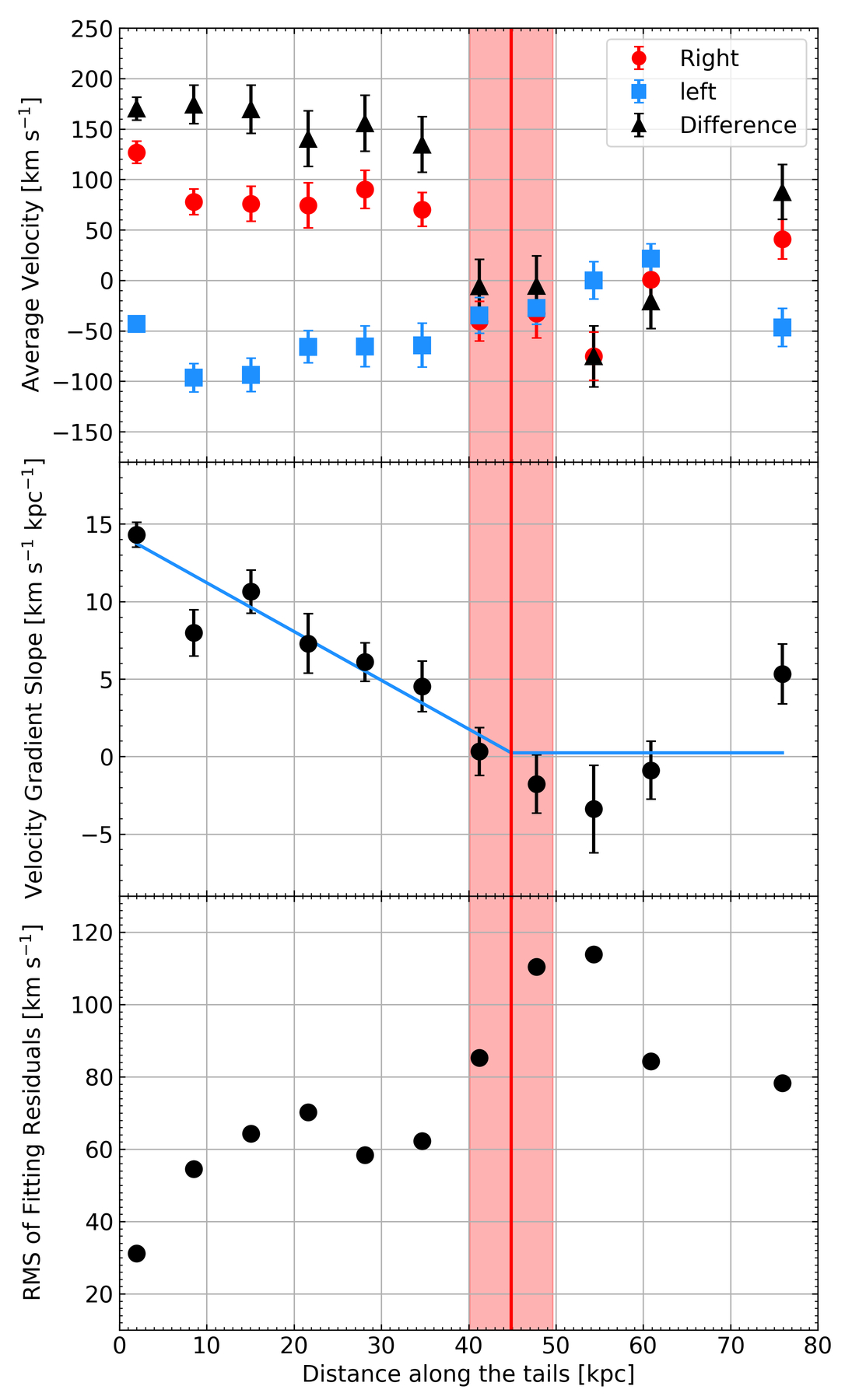}
	\vspace{-0.6cm}
	\caption{{\em Top}: The average velocities of the left and the right sides of the velocity curve 
        in each region along the stripped tails in Fig.~\ref{fig:rotation_curves_fitting} and their differences.
	{\em Middle \& Bottom}: 
	The distributions of the slope of the velocity gradient and the RMS of the fitting residuals 
        in Fig.~\ref{fig:rotation_curves_fitting} as a function of the distance along the stripped tails. 
        The blue line shows the best-fitting distribution. The end point of the velocity gradient and its 
        error are indicated by the red line and shaded red regions. We consider the red region (distance $\sim$ 45 kpc) 
        where the rotation pattern in the stripped warm, ionized gas has begun to disappear completely and the velocity 
        field is fully at the mercy of turbulence.}
	\label{fig:rotation_curves_analysis}
\end{figure}

\subsubsection{Modelling based on the velocity gradients in different regions along the stripped tails}
\label{subsubsec:velocity model:method two}
The above modeling can not trace the variation of velocity gradient along the stripped tails of ESO~137-001, which 
has been observed in Section \ref{subsec:overall properties}. Thus, we attempted the second method to construct 
the velocity field model of the ionized gas. We first rotated $48\degr$ anticlockwise for the velocity 
field to ensure the global velocity gradient along the horizontal direction. Then we divided the velocity field 
into eleven regions along the stripped tails and measured the velocity gradients in these regions. We adopted the 
same width for the first ten regions, while the width of the last region was set to be larger to cover enough 
data points. As shown in Fig.~\ref{fig:rotation_curves_fitting}, for each region, we calculated the H$\alpha$ 
flux-weighted averages of the velocities from the spaxels in each column. The corresponding velocity gradient 
was derived by fitting these average velocities as a function of the distance across the stripped tails. 
We calculated the error-weighted averages of the data in ten distance bins and adopted a linear function 
to perform the fitting. The best-fitting velocity gradients of these regions are used to model the velocity 
field of the ionized gas. 

The velocity field model of the ionized gas and its corresponding residual velocity field are shown in 
Fig.~\ref{fig:gasvel_modelling_results_velgradient}. We also performed similar kernel density estimations as 
those in Section \ref{subsubsec:velocity model:method one} to examine the velocity field model. As presented 
in Fig.~\ref{fig:reskde_regions_vel_method2} and \ref{fig:reskde_tails_vel_method2}, the level of residual 
velocity is generally lower than that in the first method, especially in the rear half of the stripped tails, 
even the level of the remaining scatter is still comparable. In addition, the fluctuation of the median residual 
velocity along the stripped tails also becomes smaller. These results imply the second method can provide better 
modelling for the velocity field of the ionized gas. Note that the second method may potentially introduce more 
parameters in the velocity field model. A better velocity model could also be due to more regions being defined 
in the stripped tails (i.e, more parameters in the model). To avoid this effect, we constrained the number of 
regions in the above modelling to properly sample the variation of velocity gradient in the stripped tails. 
We also chose the appropriate region size to ensure enough data points are included in each region.

As shown in Fig.~\ref{fig:rotation_curves_fitting}, the velocity gradient of the ionized gas changes along 
the stripped tails. In the first six regions, the velocity gradients are clearly shown and present a decreasing 
trend. While in the last five regions, the velocity gradients become less significant. To further quantify this 
variation, we plotted the slope of the velocity gradient vs. the distance along the stripped tails 
in Fig.~\ref{fig:rotation_curves_analysis}. The slope decreases as a function of the distance and 
becomes $\sim$ 0 at the last five regions, suggesting the residual rotation imprint disappears 
beyond $\sim$ 45 kpc from the galaxy. We adopted a piecewise function including a linear function and one plateau 
to fit the slope of the velocity gradient as a function of the distance along the stripped tails and obtained the 
end point of the velocity gradient at $44.9\pm4.8$ kpc. The corresponding RMS of residual velocity 
(observed velocity - velocity gradient) in the last five regions are generally larger than those in the first six 
regions, suggesting the small-scale velocity fluctuations increase in the rear half of the stripped tails. 
In the situation with strong velocity fluctuations, it could be difficult to identify the clear velocity gradient. 
We also calculate the average velocities of the left and the right sides of the velocity curve in each region along 
the stripped tails in Fig.~\ref{fig:rotation_curves_fitting} and their differences. Again, the difference nearly 
disappears beyond $\sim$ 45 kpc. Thus, while the residual rotation inherited from the galaxy is observed 
to $\sim$ 45 kpc from the galaxy, its imprint disappears beyond $\sim$ 45 kpc. One may notice a potential velocity 
gradient in the last region of the stripped tails. This signature is only dominated by the small sections of data 
points at the ends of the central and southern tails. The trend of the velocity gradient increasing again is not clear.

\section{Discussion}
\label{sec:discussion}
\subsection{The current state of the RPS tails}
RPS has been considered as an outside-in process (see the review by \citealt{Boselli2022}). 
The less-denser ISM in the outer galaxy disc is more easily and early stripped by the ram pressure than 
the denser one in the inner galaxy disc. Thus, the medium stripped from different parts of the galaxy 
is at different evolutionary stages. The overall distribution and kinematics of the ionized gas in 
ESO~137-001 and its stripped tails support the above scenario, which has also been discussed 
by \citet{Fumagalli2014} and \citet{Jachym2019}. The width of the northern and southern tails is much 
larger with respect to the distribution of the ionized gas in the galaxy disc. In addition, the velocity 
of the ionized gas in these tails is generally more blueshifted or redshifted than that in the central tail. 
These observational signatures suggest an outer-disc origin for the northern and southern tails. 
The characteristic of the ionized gas in the galaxy disc indicates the gas content has been fully peeled 
off from the outer disc during the RPS process, while the ICM wind is still stripping the ionized gas 
from the inner galaxy disc to feed the central tail.

There are clear separations between the three stripped tails of ESO 137-001, which may imply the distribution 
of ISM in the galaxy is not continuous and one possibility involves the existence of an inner gas ring. For a barred 
galaxy, an inner ring structure is often observed at the corotation resonance and encircling the bar, with the 
active star formation and concentration of the ionized and cold gas (e.g, \citealt{Combes1985,Buta1996,Comeron2014}). 
The morphology of ESO 137-001 is discussed in \citet{Waldron2022} with the {\em HST} data. There is no strong 
evidence for the existence of a bar but the inner morphology of the galaxy may be largely distorted by dust. 
The spilt of stripped tails may relate to the detailed ISM distributions in galaxies \citep{Sun2010,Jachym2014,Jachym2019}. 
In addition, other mechanisms may also contribute to the spilt of stripped tails, such as the effect of magnetic 
fields in producing the filamentary morphology of tails (e.g., \citealt{Ruszkowski2014}).

Based on the IFS observations of the front part of the primary tail, \citet{Fumagalli2014} found the velocity 
dispersion of the ionized gas also shows an enhancement to $\sim20$ kpc downstream and it can reach a peak value 
greater than 100 km s$^{-1}$ beyond this distance. Our observations confirm this signature in the primary tail 
and present similar behaviours in the northern and southern tails, which show an increased velocity dispersion 
of the ionized gas from $\sim12$ kpc to $\sim20$ kpc downstream (see the velocity dispersion map in 
Fig.~\ref{fig:2d_maps_starPA}). The typical value of velocity dispersion increases from $\sim35$ km s$^{-1}$ 
to $\sim80$ km s$^{-1}$ and maintains the same level in the remaining part of the stripped tails. These results 
imply an enhancement of the degree of turbulence in the stripped tails, which could be due to the oscillations 
caused by the fallback gas flows in the galaxy wake. As shown in the hydrodynamic simulations of RPS tails, 
the interactions between the stripped tails and ambient ICM drive the gas moving into the shadow of the galaxy 
and induce the fallback gas flows (e.g., \citealt{Roediger2007,Roediger2008,Tonnesen2009,Roediger2015,Tonnesen2019}). 
The mixing between the fallback and forward gas flows produces oscillations in the striped tails and may increase 
the degree of turbulence. The analysis based on the velocity structure function will help to further understand 
the origin and evolution of turbulence in the stripped tails (e.g., \citealt{Li2022}).

\subsection{The origin and implication of the velocity gradient decreasing in the stripped tails}
The observations of the extraplanar gas in edge-on galaxies have shown its rotation velocity decreases 
along the vertical direction of the midplane of the galaxy 
(e.g, \citealt{Heald2006,Oosterloo2007,Bizyaev2017,Levy2019,Marasco2019,Rautio2022}).
This velocity lag is considered as an important signature related to the origins of the extraplanar gas, 
which include the internal one - ejected through galaxy fountains (e.g, \citealt{Shapiro1976,Bregman1980}),
the external one - accretions from the IGM into the halo (e.g, \citealt{Binney2005,Kaufmann2006}), and a combination 
of them (e.g, \citealt{Fraternali2008}). For the extraplanar ionized gas, \citet{Levy2019} measured a median velocity 
lag of 21 km s$^{-1}$ kpc$^{-1}$ from a sample of 25 edge-on galaxies and also derived a median value 
of 25 km s$^{-1}$ kpc$^{-1}$ from the literature on the lag measurements. Based on the stellar rotation curve 
and the first velocity gradient in Fig.~\ref{fig:rotation_curves_fitting}, we assumed an initial rotation velocity 
of $\sim100$ km s$^{-1}$ of the stripped gas. Considering this velocity becomes zero at $\sim45$ kpc downstream 
(i.e., the place where the velocity gradient disappears), we thus derived a "velocity lag" 
as $\sim2$ km s$^{-1}$ kpc$^{-1}$, which suggests the velocity gradient decrease in the stripped tails is much 
slower than the velocity lag of the extraplanar gas. Since ESO~137-001 (thus the stripped gas) is moving much faster 
than the extraplanar gas, the relatively slow decrease of the velocity gradient in the stripped tails is not surprising. 
Another possible reason is that the stripped gas could be denser than the extraplanar gas. It is obvious that the place 
where the velocity gradient disappears is beyond the halo region of ESO~137-001. Thus, the effect of velocity lag in the 
galactic halo is not enough to account for the fading behaviour of the velocity gradient in the stripped tails. 
The velocity gradient itself introduces a velocity difference from that of the surrounding ICM, 
which can cause a local ram pressure to reduce the rotation motion of the stripped gas. This is probably another 
mechanism to explain the velocity gradient decrease in the stripped tails. Based on the high spectral resolution 
Fabry-Perot data, a vertical decrease in the rotation velocity of the stripped gas has also been discovered in 
the RPS galaxy NGC 4330, which is consistent with the hydrodynamic simulation of a face-on stripping process 
in this galaxy \citep{Sardaneta2022}.

From the velocity gradient decrease in the stripped tails of ESO~137-001, we can infer the kinematic evolution 
of the mixing between the stripped ISM and its ambient ICM. When the ISM is just stripped out from the galaxy, it can 
partially maintain the imprint of the galactic rotation, which is shown as the velocity gradient in the stripped tails. 
With the stripped gas moving away from the galaxy, its interactions with the ICM will gradually clean this imprint 
of the galactic rotation and make its kinematic behaviour similar to that of the ICM. As shown in the tails of ESO~137-001, 
the velocity gradient disappears beyond $\sim45$ kpc downstream, where the stripped gas becomes co-moving with the ICM 
and kinematically mixed with it. For the stripped gas, we can estimate a mixing time-scale of $\sim90$ Myr by assuming 
an average gas velocity of $\sim500$ km s$^{-1}$ along the tails (e.g, \citealt{Tonnesen2010}). 

\cite{Gronke18} gave a characteristic radius for the embedded cold clouds to survive via growth from cooling, $r_{\rm crit} \approx$ 20 pc $T_{\rm cl, 4}^{5/2}$ $\mathcal{M}$ $P_{\rm ICM, 4}^{-1}$ $\Lambda_{\rm mix, -21.4}^{-1}$ ($\chi/10^{4}$), where $T_{\rm cl, 4}$ is the cloud temperature in the unit of 10$^{4}$ K, $\mathcal{M}$ is the Mach number of the cloud, $P_{\rm ICM, 4}$ is the ICM pressure in the unit of 10$^{4}$ K cm$^{-3}$, $\Lambda_{\rm mix, -21.4}$ is the cooling function in the unit of 10$^{-21.4}$ erg cm$^{3}$ sec$^{-1}$, and $\chi$ is the density contrast between the cloud and the ICM. For the stripped clouds behind ESO~137-001, we assume $T_{\rm cl, 4}$=1, $n_{\rm e, ICM}$ = 10$^{-3}$ cm$^{-3}$, $T_{\rm ICM}$ = 7$\times10^{7}$ K, $\Lambda_{\rm mix}$ = 10$^{-21.8}$ erg cm$^{3}$ sec$^{-1}$ which is roughly the cooling function at $T_{\rm mix} \approx \sqrt{T_{\rm ICM} T_{\rm cl}}$ = 8$\times10^{5}$ K, and $\chi$ = $T_{\rm ICM}$/$T_{\rm cl}$ = 7$\times10^{3}$.
Thus, $r_{\rm crit} \approx$ 2.6 $\mathcal{M}$ pc. We note that this scale is very similar to the characteristic scale of the H$\alpha$ emitting gas estimated by \cite{Sun2021}.
Stripped clouds are subject to local ram pressure. The drag time $t_{\rm drag} \sim \chi r / \upsilon_{\rm cl}$, while $\upsilon_{\rm cl}$ is the velocity of the cloud relative to the local surrounding medium and $r$ is the radius the cloud. Assuming the relative velocity between the stripped cloud and the galaxy is $\upsilon_{\rm rel}$, the {\em MUSE} results suggest that $t_{\rm drag} \upsilon_{\rm rel} \approx$ 45 kpc. We can write $t_{\rm drag} \upsilon_{\rm rel}$ = 42 kpc ($\chi$ / 7$\times10^{3}$) ($r$ / 2 pc) ($\frac{\upsilon_{\rm rel}/\upsilon_{\rm cl}}{3}$), while the last ratio can be understood if $\upsilon_{\rm rel} \sim$ 500 km s$^{-1}$ as discussed above and $\upsilon_{\rm cl} \sim$ 150 km s$^{-1}$ from the galactic rotation at $>$ 8 kpc from the nucleus. Note that this estimate considered only ram pressure whereas mass transfer via cooling leads generally to a faster entrainment.
More detailed modeling and comparison with simulations are required in the future to better understand the change of the velocity gradient observed in stripped tails. 

\subsection{The connection of warm, ionized gas with gas at other phases}
\begin{figure}
	\centering
	\includegraphics[width=\linewidth]{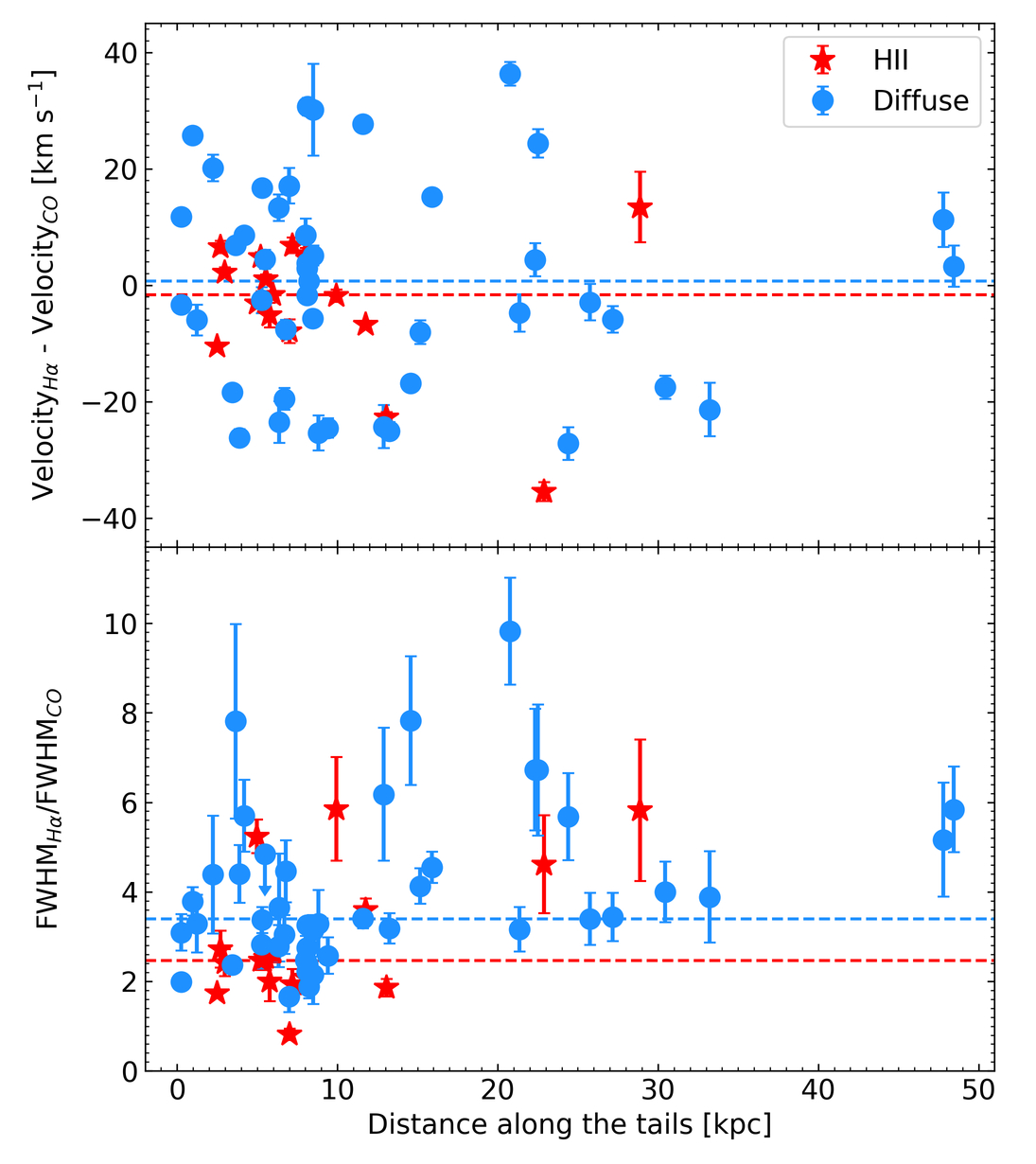}
	\vspace{-0.6cm}
	\caption{For 68 regions in ESO 137-001’s tail detected in both CO from {\em ALMA} and 
	    optical emission lines from {\em MUSE}, we compare their velocities (top panel) and velocity dispersion (bottom panel).
	    The velocity precision of {\em MUSE} is 2.5-4.0 km s$^{-1}$ \citep{Weilbacher2020}, while {\em ALMA} has a much higher 
        precision \citep{Jachym2019}. The generally small velocity difference between the two gas phases suggests they are mostly 
        co-moving through the tails. The larger velocity dispersion for the warm, ionized gas than that of the cold, molecular 
        gas can be caused by projection.}
	\label{fig:co_ha}
\end{figure}

Based on the \textit{ALMA} observations of CO(2-1) emission, the cold molecular gas has been detected 
to $\sim60$ kpc along the stripped tails of ESO~137-001 \citep{Jachym2019}. To explore the kinematics 
connection between the warm ionized gas and cold molecular gas in the stripped tails, we selected 68 
regions detected in both CO from \textit{ALMA} and optical emission lines from \textit{MUSE}, and compared 
the velocity and velocity dispersion between these two gas phases. Note our comparison involves the kinematic 
quantities along the LOS. Since the warm, ionized gas has a much larger covering fraction than the cold molecular 
gas, the region selection is limited by the {\em ALMA} data at regions with significant CO detection and a single 
velocity. The velocity and the FWHM (or 2.355$\sigma$ for a Gaussian profile) of the gas content are measured from 
the integrated spectra of individual regions. As shown in Fig.~\ref{fig:co_ha}, the velocities of the two gas phases 
generally agree within 30 km s$^{-1}$, with the RMS of the velocity difference of $\sim16$ km s$^{-1}$. In addition, 
this velocity difference is not related to the specific location of the gas content. These results suggest that the 
ionized gas and molecular gas are not only co-spatial but also co-moving through the tails, which provides further 
kinematic evidence of the mixing between different gas phases in the stripping process. As discussed in 
\citet{Jachym2019}, a part of the molecular gas is formed in situ, thus its velocity is naturally coincident with 
other gas phases. For the part of molecular gas stripped from the galaxy disc, the velocity coincidence with the 
ionized gas may also imply that two gas phases are stripped around the same time. The FWHM of ionized gas 
is typically higher than that of molecular gas, with a median ratio of $\sim3.3$. Such kind of difference in 
velocity dispersion has also been observed in the gas filaments of the cool core clusters 
(e.g, \citealt{Tremblay2018,Olivares2019}). The possible reason is due to the projection effect of different 
spatial distributions between the different gas phases (e.g., each LOS catches many tiny ionized gas clouds 
but only one big molecular cloud). \citet{Gaspari2018} shows that warm ionized gas is likely to be more 
turbulent than cold molecular gas in the filaments. \citet{Li2022} compared the velocity structure functions 
of the ionized gas and molecular gas in the stripped tails of ESO 137-001 and reached a similar conclusion.

By separating the gas content associated with the \hii{} regions and diffuse regions, we also examine if the 
above comparisons are different between these two kinds of regions. The velocity differences between the ionized 
gas and molecular gas have similar median values (-1.6 km s$^{-1}$ in \hii{} regions vs. 0.7 km s$^{-1}$ in diffuse 
regions), but different scatter (RMS: 11.8 km s$^{-1}$ in \hii{} regions vs. 17.4 km s$^{-1}$ in diffuse regions). 
The general small velocity difference suggests the co-moving of two gas phases in these regions. For the diffuse 
regions, the velocity dispersion ratio between the ionized gas and molecular gas is larger than that in the \hii{} 
regions (median values: 3.4 vs. 2.5), which may be related to projection and the more turbulent environment 
of the diffuse gas.

\section{Conclusions}
\label{sec:summary}
By combining our new \textit{MUSE} observations and the archived data, we construct a large mosaic to cover 
ESO~137-001 and its RPS tails. We studied the distribution and kinematic properties of the ionized gas and stars, 
and linked them with the RPS process in ESO~137-001 as well as the substructures of the stripped tails. The main 
results are summarised as follows.

The stripped ionized gas of ESO~137-001 splits into three tails. While the central and southern tails can extend 
to at least $\sim87$ kpc and $\sim75$ from the galaxy centre and present good continuity, the northern tail is 
more clumpy and contains several filament substructures, which cover a region of $\sim5$ kpc $\times$ 40 kpc.
Based on the kernel density estimation, we studied the distributions of the gas velocity and velocity dispersion 
along the stripped tails. The median velocity of the ionized gas does not significantly change along the stripped 
tails, which confirms the galaxy is mainly moving on the sky plane. While the median velocity dispersion of the 
ionized gas increases from $\sim35$ km s$^{-1}$ in the galaxy disc to $\sim80$ km s$^{-1}$ at $\sim20$ kpc 
downstream and maintains a similar level in the rest of the stripped tails, suggesting an enhancement of 
the degree of the turbulence within this distance. The spreads of the velocity and velocity dispersion in 
the stripped tails are generally larger than those in the galaxy disc, which is consistent with the predictions 
in the simulations of RPS tails.
 
In the galaxy disc of ESO~137-001, significant differences are shown between the distributions and velocity 
fields of the ionized gas and stars, indicating the intense perturbations produced by RPS and this process 
is already well advanced. While the stars are well located in a rotating disc, most ionized gas in the outer 
galaxy disc has been stripped. Ram pressure shapes the remnant ionized gas into a narrow cone, which connects 
the inner galaxy disc with the central tail. The velocity differences between the ionized gas and stars present 
a clear enhancement from $\sim0$ km s$^{-1}$ in the inner galaxy disc to $\sim40$ km s$^{-1}$
in the outer galaxy disc, which provides a potential hint of gas deceleration due to RPS.
 
The velocity field of the ionized gas presents a clear velocity gradient roughly perpendicular to the stripping 
direction, which ranges from an average value of $\sim-80$ km s$^{-1}$ in the northern tail to $\sim120$ km s$^{-1}$ 
in the southern tail. This velocity gradient linearly decreases along the stripped tails and disappears 
at $44.9\pm4.8$ kpc downstream. The velocity range of this gradient is generally consistent with that of the 
stellar rotation curve, indicating it originates from the rotation motion of the galaxy disc. We constructed 
the velocity model for the above residual galactic rotation based on: (1) the global velocity gradient in the 
front part of the stripped tails; (2) the velocity gradients in different regions along the stripped tails. 
The second model provides less residuals after subtracting the observed velocity field, suggesting it is better 
than the first one.

The observed results of ESO~137-001 and its stripped tails support the outside-in scenario of RPS. 
The ionized gas in the outer disc has been fully stripped to form the northern and southern tails, while the 
central tail is built with the ionized gas from the inner galaxy disc, which is still feeding the stripping 
process. By comparing with the simulated RPS tails, we interpreted that the enhanced velocity dispersion in 
the stripped tails of ESO~137-001 could be a result of the oscillations induced by the fallback gas flows 
in the galaxy wake. We discuss the possible origins of the velocity gradient decreasing in the stripped tails, 
which include the rotation velocity lag within the galaxy halo and the effect of local ram pressure caused by 
the interaction with ambient ICM. The decreasing velocity gradient in the stripped tails also implies the 
kinematic evolution of the mixing between the stripped ISM and its ambient ICM. The comparison of kinematic 
properties between the warm ionized gas and cold molecular gas shows that they are co-moving through the 
stripped tails, which provides further kinematic evidence of the mixing between different gas phases in 
the stripping process.

Our research demonstrates the great potential of wide-filed optical integral-field spectroscopy in probing the 
detailed distribution and kinematic properties of ionized gas (i.e., warm phase) in RPS. With the ongoing development 
of spatially resolved spectroscopy in X-ray, IR, and radio (e.g., \textit{Athena}, \textit{JWST}, \textit{SKA}, 
and \textit{ALMA}), future multi-wavelength campaigns will provide a comprehensive picture of the kinematic properties 
of the multi-phase stripped medium, which will further shed light on the role of gas kinematics in their mixing and 
evolution and provides important information for future simulations of RPS.

\section*{Acknowledgements}
We thank the referee for detailed and useful comments.
Support for this work was provided by the NSF grant 1714764 and the NASA grants 80NSSC21K0704 and 80NSSC19K1257.
P.J. acknowledges support from the project RVO:67985815, and the project LM2023059 of the Ministry of Education, Youth and Sports of the Czech Republic.
M.F. and M.F. acknowledge support from the European Research Council (ERC) under the European Union's Horizon 2020 research and innovation programme (grant agreement No 757535). Y.L. acknowledges financial support from NSF grants AST-2107735 and AST-2219686, and NASA grant 80NSSC22K0668. This research is based on observations collected at the European Southern Observatory under ESO programme 095.A-0512(A), 0104.A-0226(A), 60.A-9349(A) and 60.A-9100(G).

\section*{Data Availability}
The \textit{MUSE} raw data are available to download at the ESO Science Archive Facility\footnote{http://archive.eso.org/cms.html}.
The reduced data underlying this paper will be shared on reasonable requests to the corresponding authors.

\bibliographystyle{mnras}
\bibliography{ESO_137_001_Kinematics}

\bsp	
\label{lastpage}
\end{document}